\documentclass{article}

\usepackage{arxiv}
\usepackage{wrapfig}
\usepackage[utf8]{inputenc} 
\usepackage[T1]{fontenc}    
\usepackage{hyperref}       
\usepackage{url}            
\usepackage{booktabs}       
\usepackage{amsfonts}       
\usepackage{nicefrac}       
\usepackage{microtype}      
\usepackage{lipsum}		
\usepackage{graphicx}
\usepackage{natbib}
\usepackage{doi}
\usepackage{xcolor}

\title{Socio-Technological Challenges and Opportunities: Paths Forward}

\author{ 
	{Carole-Jean Wu} \\
	Facebook \\
	\And
	{Srilatha Manne}\\
	Facebook*$\diamond$ \\
    \And
    {Parthasarathy Ranganathan} \\
	Google \\
	\And
    {Sarah Bird} \\
	Microsoft \\
	\And
    {Shane Greenstein} \\
	Harvard University \\
}

\date{}

\begin{document}
\maketitle

\begin{abstract}
Advancements in digital technologies have a bootstrapping effect. The past fifty years of technological innovations from the computer architecture community have brought innovations and orders-of-magnitude efficiency improvements that engender use cases that were not previously possible -- stimulating novel application domains and increasing uses and deployments at an ever-faster pace. Consequently, computing technologies have fueled significant economic growth, creating education opportunities, enabling access to a wider and more diverse spectrum of information, and, at the same time, connecting people of differing needs in the world together. Technology must be offered that is inclusive of the world’s physical, cultural, and economic diversity, and which is manufactured, used, and recycled with environmental sustainability at the forefront. For the next decades to come, we envision significant cross-disciplinary efforts to build a circular development cycle by placing pervasive connectivity, sustainability, and demographic inclusion at the design forefront in order to sustain and expand the benefits of a technologically rich society. We hope this work will inspire our computing community to take broader and more holistic approaches when developing technological solutions to serve people from different parts of the world.

\vspace{0.25cm}

\thanks{This article is intended to capture the ISCA panel on the Microprocessor 50: Societal Challenges (see \url{https://www.iscaconf.org/isca2021/program/}) from the lens of computer architects and the following discussions. This work represents the opinions of the authors and does not reflect the position of their respective companies. 
}

\thanks{
*$\diamond$ Will be with Facebook; work started at Microsoft. 
}

\vspace{0.25cm}

\end{abstract}

\vspace{0.25cm}
{\noindent \Large \bf Introduction}

Digital technologies have had an undeniable influence on humanity’s well-being, transforming all aspects of our lives. Underpinned by advances in process technology, computer architecture, software engineering, and artificial intelligence (AI), the rapid technological development  of  the past five decades has altered the way we learn, work, commute, shop, socialize, eat, relax, and even sleep, both directly and indirectly. At the personal level, every US household has an average of 25 connected devices such as cell phones, tablets, laptops, gaming  consoles, wireless headphones, smart TVs, smart speakers, fitness trackers, and connected fitness machines~\citep{Devices}. Digital technologies have also impacted major aspects of services that we regularly use and rely upon.  Amazon warehouses are equipped with over 200,000 robots in 2020 to boost operational efficiency~\citep{Ackerman-Amazon-robots}.  AI-powered robots are a growing presence in the farming industry~\citep{robot-farms}. Medicine has been transformed by technological advances leading to the decoding of the humane genome, resulting  in genetically targeted therapies that help cancer patients survive longer or even enter full remission~\citep{HumanGenome}. Looking ahead, AI is showing great promise in solving the grand challenge in biology -- the protein structure prediction problem -- which can once again lead to revolutionary changes in the field of biological sciences ~\citep{AlphaFold}.

Technology has also aided under-privileged and vulnerable groups in surprising ways. As an example, cell phones empower women in vulnerable situations to stay connected with the world, receive education and news, and establish businesses to support their families~\citep{India-smartphone}. Emerging technologies such as surveillance cameras enable authorities to respond to violence and curb crimes. Drones deliver life-saving medical supplies in rescues~\citep{medical-drones} and robots are employed in field discovery where the environment is unsafe for humans~\citep{search-rescue}. The most recent example is the pandemic where digital technologies empowered society to stay connected and function effectively, and aided in disease tracking and drug discovery to limit the spread of the pandemic~\citep{covid-tracing,covid-digital-tracing,ai-covid-tracing}. 

Advancements in digital technologies have a bootstrapping effect. The past fifty years of technological innovations from the computer architecture community have brought innovations and orders-of-magnitude efficiency improvements that engender use cases that were not previously possible -- stimulating novel application domains and increasing uses and deployments at an ever-faster pace. Consequently, computing technologies have fueled significant economic growth, creating education opportunities, enabling access to a wider and more diverse spectrum of information, and connecting people of differing needs in the world together. 

\vspace{0.5cm}
{\noindent \Large \bf Microprocessors at 50} 

Digital technologies have witnessed significant advancement over the past five decades. The first commercially-produced microprocessor -- Intel 4004 -- was manufactured in 10,000 nm process technology in 1971, and ran at 740kHZ with 2,250 transistors~\citep{Intel-4004}. Fifty years later, the typical microprocessor is manufactured in a 5+ nm process technology and is capable of running at 5,000,000kHz (e.g.,~\citep{intel-i9,amd-ryzen}) with more than 3.9 billion transistors. This is a more than 6,750 fold improvement in processor clock speed and 1.7 million times more transistors for microprocessors manufactured in 1971 than that in 2021.

Moore’s law scaling underpins the evolution of microprocessors~\citep{moore-law}. The steady doubling of transistor density enables miniaturization of computing systems, from large mainframes to personal computers and from mobile/smartphones to Internet of Things (IoTs) and AR/VR wearables. The \textit{1990s} were the golden age of microprocessor innovations. Microarchitectural optimizations enabled impressive ILP scaling: most notably, 
in-order vs. out-of-order execution~\citep{io-ooo-2,io-ooo-3,io-ooo-1}, 
branch predictors~\citep{branch-predictor-3,branch-predictor-2,branch-predictor-1,branch-predictor-4,perceptron}, caches~\citep{cache,direct-map-cache,multi-level-caches,cache-prefetch-buffer}, prefetchers~\citep{prefetching-1,prefetching-2,MC-prefetching}, 
single vs. simultaneous multithreading~\citep{smt-1,smt-2,MC-multithreading}.

\begin{wrapfigure}{l}{0.45\textwidth}
  \vspace{-0.3cm}
  \begin{center}
  \includegraphics[width=0.4\columnwidth]{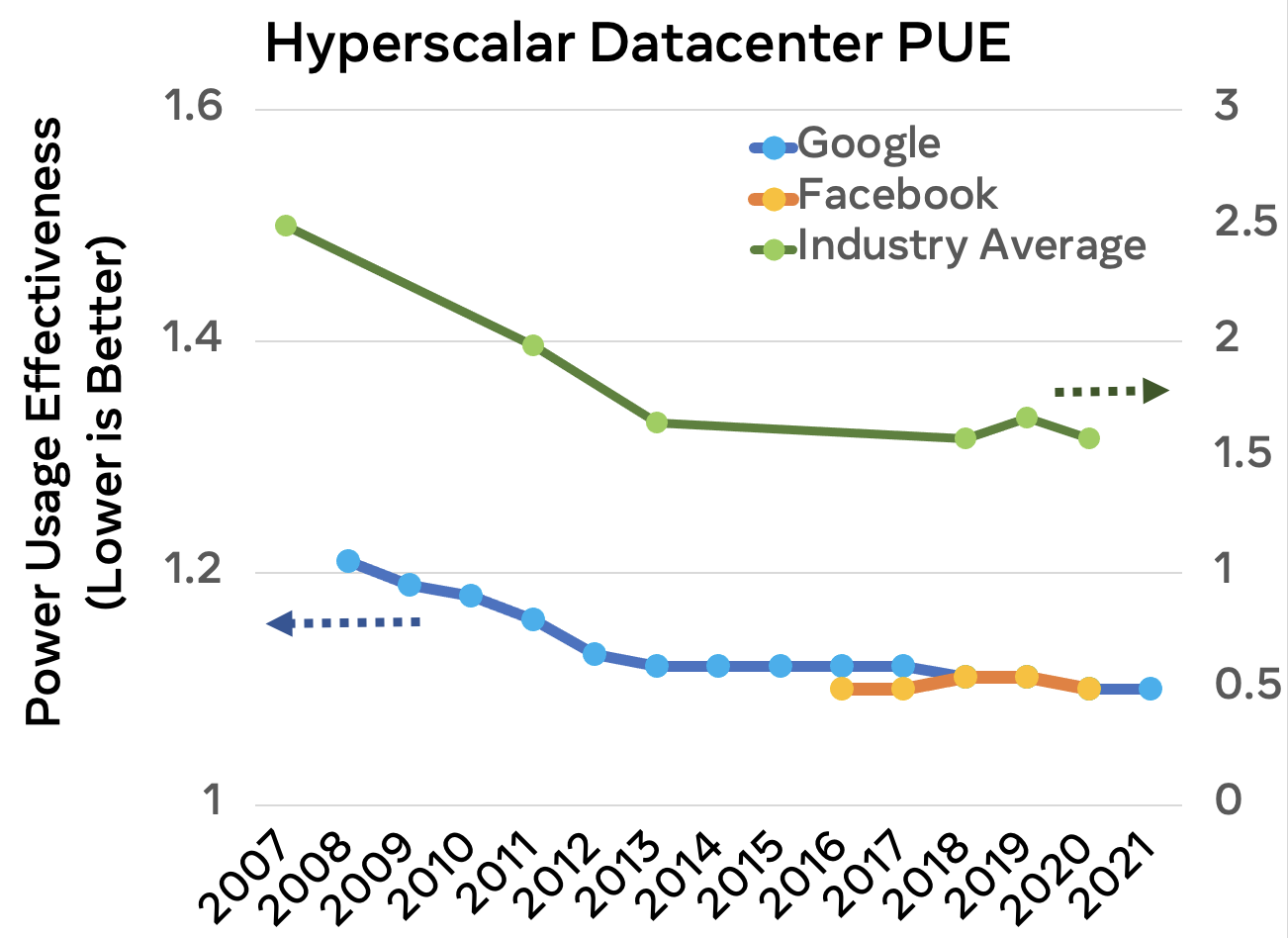}
  \end{center}
  \caption{PUE of hyperscalar datacenters, such as Google's, has improved from 1.21 (2008) to 1.10 (2021)~\citep{datacenter-google-pue} whereas the PUE of Facebook datacenters is 1.10 (2020)~\citep{datacenter-facebook-pue} and the average PUE for a typical data center in 2020 is 1.58~\citep{datacenter-pue-uptime-1,datacenter-pue-uptime}.}
  \label{fig:datacenter-pue}
\end{wrapfigure}

In the \textit{2000s}, microprocessors faced two significant challenges: the \textit{memory wall}~\citep{memory-wall} and the \textit{power wall}~\citep{power-wall-1,power-wall-2}. While processor frequencies improved with Moore’s law scaling, memory latency did not and memory subsystems increasingly gated performance. Furthermore, Dennard scaling came to an end and fine-grained, high power density thermal hot spots limited the performance of microprocessors. The Memory and Power walls subsequently drove decades of innovations in multi-core scaling~\citep{MC-cmp}, 
memory consistency and cache coherence~\citep{consistency-2,consistency-1,coherence-consistency}, cache and memory hierarchy optimization~\citep{dip,rrip,ship,MC-caches,MC-compression,MC-memory-1,MC-cache-1}, 
network-on-chip design and optimization~\citep{noc-1,noc-2,MC-noc-1,MC-noc-2},
and power- and thermal-aware design and management~\citep{wattch,dtm,temperature-microarchitecture,MC-power-efficiency,MC-power-efficiency-v1}.

During the same period, computations were migrating from client/personal devices to the cloud, demanding significant investment in large-scale data centers~\citep{datacenter-computer-1}. The location of these data centers is dictated by a myriad of constraints -- maximizing power and operational efficiency, proximity to population centers, weather conditions, local tax breaks -- leading to some interesting tradeoffs in size, location, and ownership (on premise or cloud based) of the data centers. In 2019, urban data centers that optimize for service latency responsiveness are 26.7\% smaller in size than the average data center operated by the major cloud providers~\citep{Greenstein2020WhereTC}. Furthermore, between traditional and highly optimized hyperscale data centers, power usage effectiveness (PUE) has a stark difference -- more than 40\% higher efficiency for hyperscale data centers (Figure~\ref{fig:datacenter-pue}). Going forward, the demand on fast(er) service response and availability is playing an increasingly significant role on data center site selection and computing infrastructures connecting the edge and the cloud.


The \textit{2010s} is the golden age of domain-specific architectures and specialized hardware, fueled by the rise of big data and AI~\citep{dark-silicon,dsa}. The massive economic growth opportunities of AI have revolutionized  the entire system stack design, resulting in hardware tailored to  machine learning execution ~\citep{msft-dnn-accelerator,Eyeriss,tpu,catapult,simba,mlperf-1,DL-accelerators-3,mlperf-4,facebook,DL-accelerators-1,DL-accelerators-2,gpu}from megawatt data center-scale infrastructures, to tens of watts inference engines, to micro-watt microcontrollers at the edge. Building on top of domain-specific characteristics, further system efficiency can be extracted with a rich array of application-specific accelerators at the cloud scale~\citep{cloud-accelerators-1,cloud-accelerators-2}.

\begin{wrapfigure}{r}{0.5\textwidth}
  \vspace{-0.5cm}
  \begin{center}
  \includegraphics[width=0.45\columnwidth]{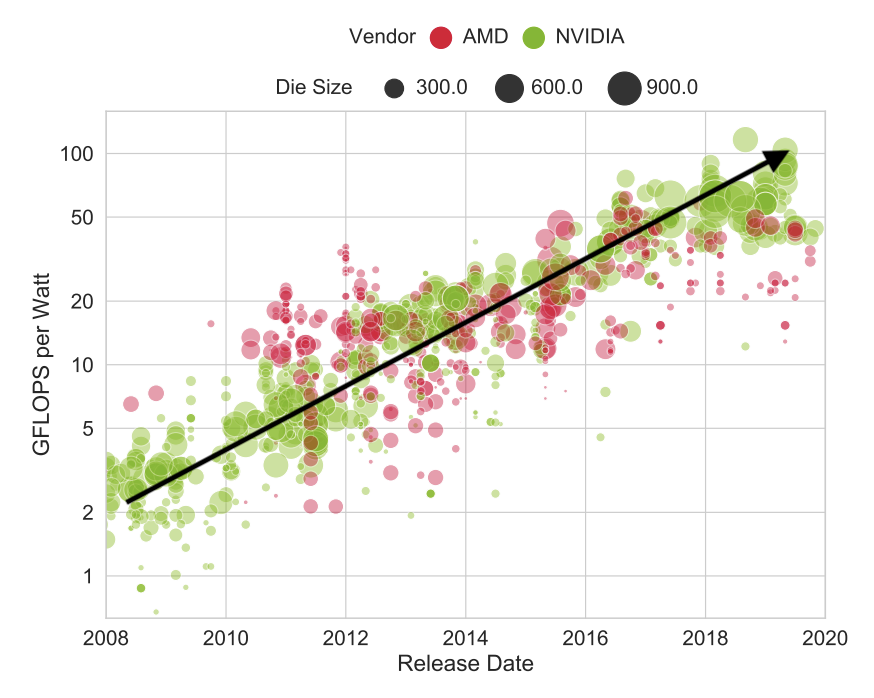}
  \end{center}
  \vspace{-0.5cm}
  \caption{As a result of Moore's law scaling and architectural optimization, GPU theoretical performance (GFLOPs) per watt doubles every 3-4 years ~\citep{cpu-gpu-ppw}.}
  \label{fig:flops-per-watt}
\end{wrapfigure}

The last fifty years of digital products have been driven by a combination of market innovation and user needs. In some cases, scientific curiosity, coupled with a market need for specialty medical drugs drove innovations such as the pursuit of the human genome. In other cases, innovations created a market such as the case for smartphones. The availability of smartphones led to further innovations and massive disruptions via  \textit{sharing economy} companies such as Uber and AirBnB~\citep{sharing-economy}. None of this would have been feasible without fundamental innovation in process technology, hardware and software design, and a laser focus on efficiency (Figure~\ref{fig:datacenter-pue}), optimization (Figure~\ref{fig:flops-per-watt}), and cost reduction. Altogether, digital technology advancement has led to \textit{efficiencies of scale} propelling decades of economic growth. 

The narrow focus on marketability, efficiency, and disruptive innovation has also resulted in  many challenging, and sometimes unexpected, societal issues. Examples include widening disparity and inequity of access to digital technologies, spread of disinformation and misinformation in online platforms, privacy and security violations at both the personal and political level, and propagation of human bias into AI training and use cases at-scale~\citep{societal-challenge-10,societal-challenge-11,societal-challenge-12,societal-challenge-13,ai-now-2018,societal-challenge-8,societal-challenge-7,societal-challenge-6,societal-challenge-4,societal-challenge-8a,societal-challenge-5,ai-now-2019,societal-challenge-2,societal-challenge-9,societal-challenge-1,societal-challenge-3}. In addition, the information and computing technology sectors consume a significant amount of global electricity, water, and natural resources, leading to paramount carbon and environmental footprint~\citep{ict-footprint-5,ict-footprint,ict-footprint-2,ict-footprint-3,ict-footprint-4,ict-footprint-1}. Digital technology is an integral part of human existence. As the field matures and as the world faces dire challenges from climate change to societal and political upheavals, a deliberate approach is required to technological innovation that centers on a positive societal impact while serving the needs of a market economy. 

\vspace{0.5cm}
{\noindent \Large \bf Looking to the Future}

Predicting the future is always difficult. How many of us could have predicted that the technology depicted in futuristic TV shows from five decades ago such as Star Trek would be commonplace today (cell phones, AI, voice recognition to name a few)? However, what is possible to foretell are the following: 
\begin{itemize}
    \item {\bf Pervasive Connectivity}: the internet and the technology enabling it will become even more fundamental to everyday life, requiring that information and communication technology infrastructure, just like power grids, be both secure and resilient;
    \item {\bf Sustainability}: Environmental pollution, resource depletion and climate change will dominate how products are designed, produced, consumed, and recycled. Systems must be manufactured with less planetary impact, use less energy while in operation, and produce less e-waste at the end of life; and 
    \item {\bf Demographic Inclusion}: Massive demographic upheavals resulting in an aging population in most of the world and rising population centers in Sub-Saharan Africa will influence what products are created and for whom, and where technology is designed and deployed in the future.  
\end{itemize}

{\large \bf Pervasive Connectivity.}
In 2021, approximately 65\% of the world’s population has access to the internet~\citep{internet-statistics}, and this is expected to improve moving forward. Seamless connectivity will be required for everyday activities from ordering groceries to driving your car. There are advantages to connectivity from more efficient driving and less food waste, easy access to educational resources, and the ability to work remotely. However, what this also implies is that our lives and livelihoods will be inexorably linked to the accessibility, privacy, security, and resilience of the IT infrastructure. 

Reliable internet access for the world's population is an essential requirement of pervasive connectivity. Despite the positive societal and economical impact, at-scale power delivery and networking infrastructure development has proven to be costly and  highly geographically-constrained. 
According to an analysis from the FCC, at least 18 million Americans did not have stable broadband in 2020~\citep{fcc-data}, and many who have access may not be able to afford it. According to a recent UN report, two-thirds of school age children do not have access to the internet in their home~\citep{unicef}. 
Diverse technology innovations are particularly needed to increase information accessibility and to connect people of differing needs in the world together in a resilient manner. Microsoft Airband~\citep{microsoft-airband} aims to expand broadband access in rural parts of the United States while Google Loon~\citep{google-loon} and Facebook Aquila~\citep{facebook-aquila} provide affordable high-speed internet to under-connected communities by overcoming physical barriers with innovative technology solutions. The recent success of placing internet communication satellites in low Earth orbit opens the door to tremendous opportunities for providing internet access to populations in challenging geographic locations~\citep{starlink}. While satellite communication removes the geographical and political boundaries, it can accelerate the spread of (mis)information. Thus, as technologies continue to enable pervasive connectivity at scale, we must take a deliberate approach to develop technology solutions responsibly. 


Along with pervasive connectivity comes the requirements for security and privacy given the prevalence of ransomware attacks and ongoing privacy hacks from both private groups and nation states. Many aspects of our public and private lives will be online and accessible to bad actors, and the internet and the technology it enables will continue to be susceptible without strong security and privacy measures in place from the top of the software stack to the underlying hardware running the code~\citep{meltdown,spectre}.

The IT infrastructure, just like in other critical infrastructure such as telecommunication and power delivery, will require backup and recovery mechanisms in order to minimize or eliminate downtime. These already exist in cloud data centers with examples such as redundant storage~\citep{Msft-redundant-storage} and network~\citep{Msft-redundant-network}, and backup batteries and generators. However, redundancy can require significant additional hardware and network bandwidth which inflates the cost of the overall cloud infrastructure. 

Recent climate events are pushing the issue of resiliency to the forefront. In the Texas power outages of 2021, many top-tier data centers were able to continue operations using diesel backup generators. However, there is a limit to how long backup systems can operate given the finite amount of fuel onsite and weather conditions making it difficult to transport more fuel to the data center~\citep{texasevent}. As catastrophic weather events become more common and last longer due to climate change, even the largest data centers may have trouble~\citep{doe-microgrid}. 
For instance, the latest fires in Oregon brought down major power grid infrastructure that connects the grids of California to Oregon at the same time as a heat wave resulted in a need for additional power for cooling~\citep{oregon-fire}. California narrowly escaped rolling blackouts this time, but these events will continue to increase in intensity and frequency as hot and dry conditions lengthen and strengthen the fire season, reduce hydro-electric power due to a lack of water, and increase the need for power for human comfort and survival. 

For IT infrastructures of the future, a one-size-fits-all solution may not work either for the type of data center or for the services that it provides. Large data centers that require many megawatts to operate may not be viable during a blackout with limited power and where critical operations such as hospitals are given priority. Smaller data centers may be more resilient either with backup generators or using a local micro-grid to run operations when the main grid is down. 
On the service side, some applications may be more tolerant of outages than others. Hence, with a limited power budget, application availability should be tiered during critical times. Other solutions include making applications resilient by design or more amenable to migration to a non-impacted data center given enough warning before catastrophic events occur~\citep{resilient-1,resilient-2}. 

Designing for resilience addresses issues that are not just related to resilience under climate duress but also solutions that are tolerant of different geographies, power and networking infrastructures. In order to achieve pervasive connectivity across the world, IT infrastructure must be available across developed and developing regions. In Africa, for example, it is difficult and expensive to develop a power grid that can serve such a large continent. Renewables, however, have the benefit of being portable and scalable, and are powering much of Africa’s latest power advances, with off-grid and micro-grid renewable energy solutions serving poorer and/or non-centralized communities~\citep{africa-energy}. Data centers serving these regions may also have to rely on micro-grids running on renewable energy that fluctuate in their capacity based on the weather. 
Hence, rather than designing for a resilient and consistent power grid, IT infrastructure may be better served by (co-)designing agile applications and data centers, such as~\citep{GreenSlot,carbon-aware-computing,dc-grid-coupling,Flex}, that implicitly assume power fluctuations and variability.

Achieving pervasive connectivity in the presence of resiliency, security, and privacy requirements can be challenging given the interrelated nature of the problems. For instance, current approaches to achieving resilient computing often rely on replication and redundancy which not only increases cost but also enlarges the surface of security and privacy preservation. Migrating applications to another region, which may be necessary due to power constraints, can expose data to further security threats. To retain physical security of data, applications may be hosted in limited geographical regions but this can increase application vulnerability to catastrophic events. In addition, secure and resilient computing infrastructures can often come with significant environmental implications. Hence, any computing infrastructure solutions must be cognizant of the multifaceted nature of the problems being addressed.  


{\large \bf Sustainability.}
Resource limitations, climate change, water depletion, electronic waste, ecosystem damage, and environmental racism are just a few of the topics under the larger sustainability umbrella. There is increased focus on these topics from both industrial and political institutions. All major technology companies have pledged to reduce or eliminate their carbon footprint in the next decade by reducing the environmental impact associated with manufacturing and using their products. Examples of such commitments are Facebook achieving NetZero in operational emissions in 2020 and across its value chain by 2030~\citep{facebook-sustainability},  Apple’s pledge for 100\% carbon neutral supply chain by 2030~\citep{apple-sustainability}, Microsoft’s goal of being carbon negative by 2030~\citep{msft-sustainability}, and Google’s aim of 24x7 carbon free data centers~\citep{google-sustainability}. 
In 2020, Amazon, Google, Facebook, and Microsoft were the top four technology companies that purchased significant renewable energy capacities, accounting for 30\% of the cumulative total from corporations globally~\citep{renewable-purchase}.
In addition, countries and trading zones are legislating carbon emission requirements. China has committed to be carbon free by 2060~\citep{china-climate} and the EU has committed to cut carbon emissions by 55\% by 2030~\citep{eu-climate}. 

Sustainability targets and the associated regulations will  continue to increase, and hardware must be manufactured with less planetary impact, use less energy while in operation, and produce less e-waste at the end of life~\citep{bio-degradable,bio-degradable-micro}. Existing practices such as the move to hyperscale data centers have already reduced IT’s carbon footprint by consolidating and sharing computing resources, and operating those resources more efficiently (Figure~\ref{fig:datacenter-pue}) -- AWS~\citep{aws-cloud}, Azure~\citep{msft-cloud}, Google cloud~\citep{google-cloud}, and Facebook datacenter infrastructures~\citep{facebook-cloud}. 
In fact, data center electricity consumption has slowed down significantly. The total energy consumption of the US data centers increased by about 4\% from 2010-2014, compared with the estimated 24\% increase from 2005-10 and nearly 90\% increase from 2000-05~\citep{Masanet}. Furthermore, despite the increase in the global data center energy use, the number of users benefiting from the cloud infrastructure have increased even more---from 2010-18, the global data center compute instances increased by 5.5 times with an estimated 6\% increase in the global data center energy use.

However, more can be done -- \textit{we must go beyond efficiency optimization and build a sustainable ecosystem to achieve environmentally-sustainable computing}. For instance, develop expandable hardware and software stack that facilitate significantly longer lifetimes than the current averages of less than 3 years for cell phones~\citep{phonedurability} and 4 to 5 years for servers~\citep{uptimesurvey}. 
Modular system design will enable component-level upgrades without having to decommission the system at its entirety, reducing overall electronic waste and the environmental footprint~\citep{fair-phone}. Other actions being discussed or implemented require manufacturers to make their systems more repairable, resulting in increased product lifetime~\citep{right-to-repair,exec-order}. Another option is to reduce the number of devices in an average household. In the US, for example, the average household is equipped with an average of 25 connected devices~\citep{Devices}. In many cases, smartphones might be powerful enough for the task, but a tablet or laptop is needed for a viable keyboard or a larger viewing platform. Some of these additional devices could be replaced with a virtual reality or augmented reality solution that is both portable and can provide virtual keyboards and visual clarity without requiring additional hardware.
To achieve an environmentally sustainable computing future, we will have to build a \textbf{circular economy} for computing that supports the design principle of \textit{reduce}, \textit{reuse}, \textit{repair}, and \textit{recycle}.
These and other potential solutions likely require a complete redesign of the software and hardware stacks both at the edge, within the cloud, and in the edge-cloud collaborative execution environment, in order to provide resilient, long lasting, innovative solutions. 

Making technology more sustainable is only one part of the technical challenge. There is another side to the story---there are significant sustainability benefits resulting from computing technology. Programs, such as Farm Beats~\citep{farm-beats}, address how to optimize operations on a farm using technology that is inexpensive and readily available to people in rural areas. Food production accounts for 19\% of the world’s carbon emissions, and producing food more efficiently and with less toxicity has long term benefits for the world~\citep{gates}. 
For the residential sector in the US, space heating and cooling contributed to over 40\% of the total electricity consumption~\citep{us-electricity-use}. This is where smart home IoT devices, such as Nest, can have an impact. 
AI is used to discover new electrocatalysts for more efficient and scalable ways to store and use renewable energy~\citep{open-catalyst} while also being used to predict renewable energy availability ahead of actual generation to better utilize the energy~\citep{AI-load-shaping}.
Another example is the current Covid outbreak. As horrendous as the outbreak has been and continues to be, technology has enabled a portion of the economy to continue to operate even as employees work remotely. In addition, the global carbon emissions for 2020 dropped by 6.4\% with vehicle transportation in the US accounting for a portion of the global reduction~\citep{covid-co2}. Looking forward, information technology can improve efficiencies in practically every sector, from manufacturing to food production to transportation to controlling the climate in our homes and offices. Although there is a carbon cost associated with manufacturing and operating the IT ecosystem, this cost must be evaluated holistically~\citep{green-server-design,imec-iedm,ict-footprint-4,ai-footprint} in light of the benefits such an ecosystem can provide in other domains~\citep{ai-social-good-2,ai-social-good-1,ai-environment,block-chain-footprint}. 

\begin{wrapfigure}{r}{0.5\textwidth}
  \vspace{-0.3cm}
  \begin{center}
  \includegraphics[width=0.5\columnwidth]{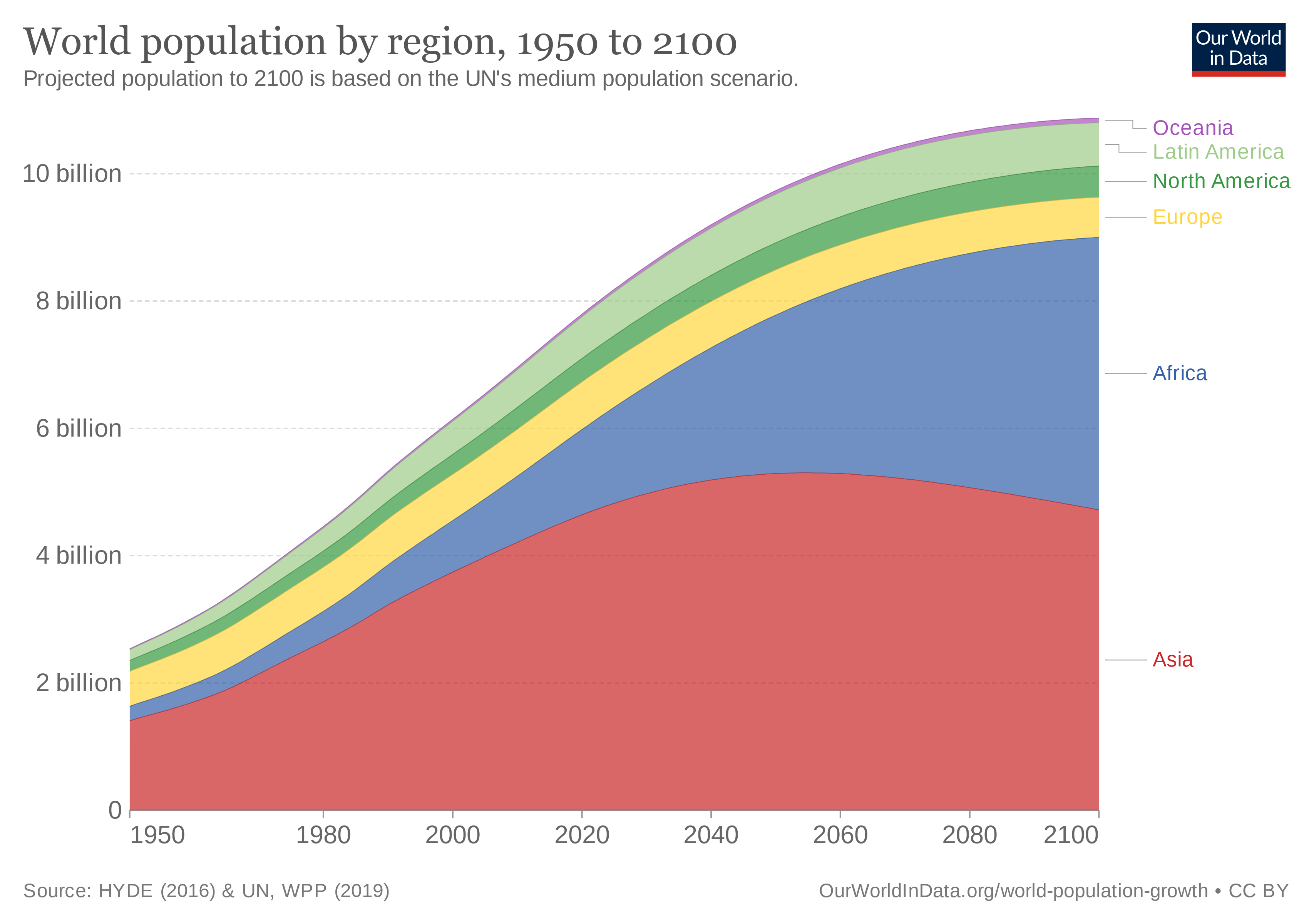}
  \end{center}
  \vspace{-0.3cm}
  \caption{Population growth estimation based on United Nations medium growth projections~\citep{UNDemographics}.}
  \label{fig:demographics}
\vspace{-0.3cm}
\end{wrapfigure}

{\large \bf Demographic Inclusion.}
Recent data has pointed to significant demographic changes that will be occurring by the year 2100~\citep{Vollset}. These include a radically declining and/or aging population in Europe, North America and Asia, and an increasing population of younger workers in Sub-Saharan Africa. Figure~\ref{fig:demographics} shows one population projection using data from the United Nations~\citep{UNDemographics}, indicating flattening or decreasing populations in most of the world except for Africa. 
The population decline is strongly correlated with an improved quality of life and educational achievement of women -- much of which is being facilitated by technologically engendered developments such as portable devices and the internet -- along with access to reproductive services. This trend holds regardless of the country, culture, or religion. These demographic changes will have a seismic impact on all societies, and will also dominate \textit{what} and \textit{where} technology is designed and deployed, respectively, in the future.

For technology to be truly inclusive in a fully-connected world, it must be available and usable for \textit{everyone} -- regardless of physical capabilities, geographic restrictions, or economic constraints. With the projected demographic changes, assistive technologies to address the needs of an aging population with physical restrictions will be essential for enhancing  an individual’s ability to be a viable and contributing member of society. 

There is a rich history of intersection between assistive technology and everyday products. Word prediction was used as an assistive technology aid for individuals with physical or developmental difficulties before becoming more mainstream~\citep{Jacobs2015TheEO}. Close captioning helps the hearing impaired, but it is also used by people with average hearing to capture the conversation in a noisy environment or to help with accents in the dialogue. The reverse is also true -- some of the most innovative technologies in the last few years are also inclusive of people with physical limitations. Smart speakers, such as Siri or Alexa, enable people with limited sight or physical mobility to communicate via verbal commands. Individuals with hearing issues can use their Apple AirPods as basic hearing aids~\citep{airpods}, and people with average hearing can use AI-enabled voice assistants to eliminate background noise and focus on the conversation at hand~\citep{nachmani2020voice}. For the sight impaired, virtual reality glasses are improving the eyesight of patients with degenerative eye disease~\citep{vr}. 

The virtual office technology and work culture developed during the Covid pandemic will continue to benefit individuals with physical limitations who may be unable to travel to work. Similarly, with the largest population of young workers residing in Africa over the next decades, innovative virtual office technology that can create seamless cohesion across geographies and cultures will be critical to economic success. Looking to the future, there is a clear demand for high-quality virtual reality experience in the first-person perspective. However, realizing smooth virtual space experience requires disruptive technology innovations -- hardware architectures beyond Von Neumann, high pixel density displays~\citep{oled,oled-1}, near/in-sensor computing~\citep{Stagioni}, flexible electronics~\citep{flexible-e}, or even brain-computer interface solutions~\citep{BCI-2,BCI}.

Another requirement of inclusive technology, especially in light of projected demographic changes, is that it must operate across different geographic boundaries. In 2021, much of modern technology assumes some degree of development and infrastructure, whether it be availability of the latest computing systems, broadband access, cell towers or power sources. However, much of the current world, including parts of developed countries, do not currently have unfettered access to these facilities. Even more daunting, approximately 770 million people, or about 1 out of 10 people in the world, do not have access to a stable supply of electricity~\citep{world-electricity}. 

In order to meet the upcoming needs of the world, technological growth and resources must focus on under-served areas. This includes developing data centers for regions that may not have a resilient power infrastructure or an infrastructure that is based on local renewable energy sources and does not tap into a regional grid. 
IT devices must be operable under adverse conditions where the local network and/or power may not be available 24/7 or varies based on external conditions~\citep{CLink,poster,intermittent-computing,demographic-inclusion-2}. 
For instance, devices need to operate for days without a recharge or must operate locally until wireless or broadband is available~\citep{delay-tolerance-prediction} to ensure delay tolerance while maintaining user experiences. 
Energy storage technology plays a crucial role to smooth the intermittent nature of renewable energy generation but the cost must be significantly improved for practical deployment~\citep{energy-grid}. 
And, for technology to be truly inclusive, the way AI technologies are developed and used must be human-centered, driven by the cultural and demographic differences in the population and with pro-social goals~\citep{recsys}. Furthermore, given its increasingly large impact on the society, AI must be developed with deliberation to construct fairer and more inclusive decisions and, at the same time, it must be adopted responsibly~\citep{openai-rai,nist-trustworthy}. AI-powered products must be transparent with users on how data is gathered, used, and stored, and with controls to disable it~\citep{ibm-rai,microsoft-rai,google-rai,facebook-rai,dod-rai}. 

Finally, any IT technology must be economically-accessible to most of the world’s population. Unfortunately, even the most basic form of computing devices, a cell phone, is prohibitively expensive for much of the world. According to the Alliance for Affordable Internet, nearly 2.5 billion people live in countries where a basic cell phone would cost nearly a quarter or more of the average monthly income~\citep{a4ai}. Making the internet available to everyone will not solve the problem if the devices commonly used to access the internet are a luxury item for the world’s poor~\citep{demographic-inclusion-1}. In addition, many of the devices currently being used are inexpensive basic phones. Hence, solutions must be designed for “dumb” phones in order to reach the poorest communities. Even if users eventually have smartphones, data rates and lack of free wifi will limit the utility of data-heavy applications. Without reasonable device and application availability for the poorest communities, reaching the goal of pervasive connectivity will be difficult if not impossible and the digital divide will continue to widen between the rich and the poor. 
And, all technological solutions come with environmental impact that will inadvertently impact the marginalized communities the most. Thus, when developing technological solutions, we must keep in mind \textit{why} products are created and \textit{for whom}, and \textit{where} technology is designed and deployed, such that we consciously build environmentally-sustainable, socially-responsible, inclusive technologies for the next decades to come.

\vspace{0.5cm}
{\noindent \Large \bf Conclusion}

Predicting the innovations of the future is a difficult if not impossible task. However, the political, societal, and environmental challenges the world faces are clear, and technology can play a significant role in helping societies adapt and thrive. The problems and solutions in the three domains of pervasive connectivity, sustainability, and inclusion are interconnected and must be addressed holistically. For instance, adapting the IT infrastructure to climate change must not in turn make climate change worse.  Making innovative devices and software to address the needs of an aging population does not mean that poorer populations in other regions can be ignored. And population decline does not fully address sustainability concerns -- as standard of living improves with the help of technology, the per-capita 
environmental footprint of the population also increases~\citep{ONeill2018AGL}. Technology must be offered that is inclusive of the world’s physical, cultural, and economic diversity, and which is manufactured, used, and recycled with environmental sustainability at the forefront. We hope to inspire our architecture community to take broader and more holistic approaches when developing technologies which include a deeper understanding of the biological, societal, cultural, environmental, political, and economic implications of future technological innovations. To summarize: 
\begin{itemize}
    \item Build resilient, secure infrastructures -- software, hardware, and everything in-between -- for the computing spectrum of systems at all sizes and scales.
    \item Design for the entire system life cycle from manufacturing to end-of-life and for the entire software development cycle from experimentation to deployment. Build environmentally-sustainable systems, software, and algorithms -- beyond energy efficiency. 
    \item Broaden the scope of technologies to serve people from different parts of the world and with differing needs by enabling equitable access to technology and the rich economic opportunities it presents.
\end{itemize}
For the next decades to come, we envision significant cross-disciplinary efforts to build a circular development cycle by placing \textit{pervasive connectivity}, \textit{sustainability}, and \textit{demographic inclusion} at the design forefront in order to sustain and expand the benefits of a technologically rich society. 

\vspace{0.5cm}
{\noindent \Large \bf Acknowledgement}

We would like to thank Doug Carmean, David Brooks, and Lizy John for their insightful feedback on this work.

\begin{flushleft}
\bibliographystyle{unsrtnat}
\bibliography{references}

\begin{thebibliography}{185}
\providecommand{\natexlab}[1]{#1}
\providecommand{\url}[1]{\texttt{#1}}
\expandafter\ifx\csname urlstyle\endcsname\relax
  \providecommand{\doi}[1]{doi: #1}\else
  \providecommand{\doi}{doi: \begingroup \urlstyle{rm}\Url}\fi

\bibitem[Deloitte(2021)]{Devices}
Deloitte.
\newblock {How the pandemic has stress-tested the crowded digital home}.
\newblock
  \url{https://www2.deloitte.com/content/dam/insights/articles/6978_TMT-Connectivity-and-mobile-trends/DI_TMT-Connectivity-and-mobile-trends.pdf},
  2021.

\bibitem[Ackerman(2021)]{Ackerman-Amazon-robots}
Evan Ackerman.
\newblock {What’s Going on With Amazon’s "High-Tech" Warehouse Robots?}
\newblock
  \url{https://spectrum.ieee.org/automaton/robotics/industrial-robots/whats-going-on-with-amazons-hightech-warehouse-robots},
  June 2021.

\bibitem[Sheikh(2020)]{robot-farms}
Knvul Sheikh.
\newblock {A Growing Presence on the Farm: Robots}.
\newblock
  \url{https://www.nytimes.com/2020/02/13/science/farm-agriculture-robots.html},
  February 2020.

\bibitem[Hum(2021)]{HumanGenome}
{How the Human Genome Project shook the world of cancer research}.
\newblock
  \url{https://www.icr.ac.uk/news-features/latest-features/how-the-human-genome-project-shook-the-world-of-cancer-research},
  February 2021.

\bibitem[Jumper et~al.(2021)Jumper, Evans, Pritzel, Green, Figurnov,
  Ronneberger, Tunyasuvunakool, Bates, Žídek, Potapenko, Bridgland, Meyer,
  Kohl, Ballard, Cowie, Romera-Paredes, Nikolov, Jain, Adler, Back, Petersen,
  Reiman, Clancy, Zielinski, Steinegger, Pacholska, Berghammer, Bodenstein,
  Silver, Vinyals, Senior, Kavukcuoglu, Kohli, and Hassabis]{AlphaFold}
John Jumper, Richard Evans, Alexander Pritzel, Tim Green, Michael Figurnov,
  Olaf Ronneberger, Kathryn Tunyasuvunakool, Russ Bates, Augustin Žídek, Anna
  Potapenko, Alex Bridgland, Clemens Meyer, Simon A.~A. Kohl, Andrew~J.
  Ballard, Andrew Cowie, Bernardino Romera-Paredes, Stanislav Nikolov, Rishub
  Jain, Jonas Adler, Trevor Back, Stig Petersen, David Reiman, Ellen Clancy,
  Michal Zielinski, Martin Steinegger, Michalina Pacholska, Tamas Berghammer,
  Sebastian Bodenstein, David Silver, Oriol Vinyals, Andrew~W. Senior, Koray
  Kavukcuoglu, Pushmeet Kohli, and Demis Hassabis.
\newblock Highly accurate protein structure prediction with alphafold.
\newblock \emph{Nature}, 2021.

\bibitem[Pathak(2021)]{India-smartphone}
Sushmita Pathak.
\newblock {India's All-Female News Outlet Battles Sexism, Caste — And Hits
  The Silver Screen}.
\newblock
  \url{https://www.npr.org/sections/goatsandsoda/2021/04/04/980097004/indias-lowest-caste-has-its-own-news-outlet-and-shes-in-charge},
  April 2021.

\bibitem[Nyaaba and Ayamga(2021)]{medical-drones}
Albert~Apotele Nyaaba and Matthew Ayamga.
\newblock Intricacies of medical drones in healthcare delivery: Implications
  for africa.
\newblock \emph{Technology in Society}, 66:\penalty0 101624, 2021.

\bibitem[Arnold et~al.(2018)Arnold, Yamaguchi, and Tanaka]{search-rescue}
Ross Arnold, Hiroyuki Yamaguchi, and Toshiyuki Tanaka.
\newblock Search and rescue with autonomous flying robots through
  behavior-based cooperative intelligence.
\newblock \emph{Journal of International Humanitarian Action}, 3, 12 2018.

\bibitem[Budd et~al.(2020)Budd, Miller, Manning, Lampos, Zhuang, Edelstein,
  Rees, Emery, Stevens, Keegan, Short, Pillay, Manley, Cox, Heymann, Johnson,
  and McKendry]{covid-tracing}
Jobie Budd, Benjamin Miller, Erin Manning, Vasileios Lampos, Mengdie Zhuang,
  Michael Edelstein, Geraint Rees, Vincent Emery, Molly Stevens, Neil Keegan,
  Michael Short, Deenan Pillay, Ed~Manley, Ingemar Cox, David Heymann, Anne
  Johnson, and Rachel McKendry.
\newblock Digital technologies in the public-health response to covid-19.
\newblock \emph{Nature Medicine}, 26:\penalty0 1--10, 08 2020.

\bibitem[Salath{\'e} et~al.(2020)Salath{\'e}, Althaus, Anderegg, Antonioli,
  Ballouz, Bugnion, {\v C}apkun, Jackson, Kim, Larus, Low, Lueks, Menges,
  Moullet, Payer, Riou, Stadler, Troncoso, Vayena, and von
  Wyl]{covid-digital-tracing}
Marcel Salath{\'e}, Christian~L. Althaus, Nanina Anderegg, Daniele Antonioli,
  Tala Ballouz, Edouard Bugnion, Srdjan {\v C}apkun, Dennis Jackson, Sang-Il
  Kim, James~R. Larus, Nicola Low, Wouter Lueks, Dominik Menges, C{\'e}dric
  Moullet, Mathias Payer, Julien Riou, Theresa Stadler, Carmela Troncoso, Effy
  Vayena, and Viktor von Wyl.
\newblock Early evidence of effectiveness of digital contact tracing for
  sars-cov-2 in switzerland.
\newblock 2020.

\bibitem[{Facebook AI}(2020)]{ai-covid-tracing}
{Facebook AI}.
\newblock {Using AI to help health experts address the COVID-19 pandemic}.
\newblock
  \url{https://ai.facebook.com/blog/using-ai-to-help-health-experts-address-the-covid-19-pandemic/},
  2020.

\bibitem[Intel()]{Intel-4004}
Intel.
\newblock {The Story of the Intel 4004}.
\newblock
  \url{https://www.intel.co.uk/content/www/uk/en/history/museum-story-of-intel-4004.html}.

\bibitem[Intel(2019)]{intel-i9}
Intel.
\newblock {Intel Core i9-9900KS Processor}.
\newblock
  \url{https://ark.intel.com/content/www/us/en/ark/products/192943/intel-core-i9-9900ks-processor-16m-cache-up-to-5-00-ghz.html},
  2019.

\bibitem[AMD(2020)]{amd-ryzen}
AMD.
\newblock {AMD Ryzen 9 5950X Desktop Processors}.
\newblock \url{https://www.amd.com/en/products/cpu/amd-ryzen-9-5950x}, 2020.

\bibitem[Moore(1965)]{moore-law}
Gordon~E. Moore.
\newblock Cramming more components onto integrated circuits.
\newblock \emph{Electronics}, 1965.

\bibitem[Tomasulo(1967)]{io-ooo-2}
R.~M. Tomasulo.
\newblock An efficient algorithm for exploiting multiple arithmetic units.
\newblock \emph{IBM Journal of Research and Development}, 11\penalty0 (1),
  1967.

\bibitem[Smith(1982{\natexlab{a}})]{io-ooo-3}
James~E. Smith.
\newblock Decoupled access/execute computer architectures.
\newblock In \emph{Proceedings of the 9th Annual Symposium on Computer
  Architecture}, page 112–119, 1982{\natexlab{a}}.

\bibitem[Hwu and Patt(1986)]{io-ooo-1}
W.~Hwu and Y.~N. Patt.
\newblock Hpsm, a high performance restricted data flow architecture having
  minimal functionality.
\newblock In \emph{Proceedings of the 13th Annual International Symposium on
  Computer Architecture}, page 297–306, 1986.

\bibitem[Smith(1981)]{branch-predictor-3}
James~E. Smith.
\newblock A study of branch prediction strategies.
\newblock In \emph{Proceedings of the 8th Annual Symposium on Computer
  Architecture}, page 135–148, 1981.

\bibitem[Lee and Smith(1984)]{branch-predictor-2}
Lee and Smith.
\newblock Branch prediction strategies and branch target buffer design.
\newblock \emph{Computer}, 17\penalty0 (1):\penalty0 6--22, 1984.

\bibitem[Pnevmatikatos et~al.(1993)Pnevmatikatos, Franklin, and
  Sohi]{branch-predictor-1}
Dionisios~N. Pnevmatikatos, Manoj Franklin, and Gurindar~S. Sohi.
\newblock Control flow prediction for dynamic ilp processors.
\newblock In \emph{Proceedings of the 26th Annual International Symposium on
  Microarchitecture}, page 153–163, 1993.

\bibitem[Yeh et~al.(1993)Yeh, Marr, and Patt]{branch-predictor-4}
Tse-Yu Yeh, Deborah~T. Marr, and Yale~N. Patt.
\newblock Increasing the instruction fetch rate via multiple branch prediction
  and a branch address cache.
\newblock In \emph{Proceedings of the 7th International Conference on
  Supercomputing}, page 67–76, 1993.

\bibitem[Jimenez and Lin(2001)]{perceptron}
D.A. Jimenez and C.~Lin.
\newblock Dynamic branch prediction with perceptrons.
\newblock In \emph{Proceedings of the Seventh International Symposium on
  High-Performance Computer Architecture}, 2001.

\bibitem[Smith(1982{\natexlab{b}})]{cache}
Alan~Jay Smith.
\newblock Cache memories.
\newblock \emph{ACM Computing Surveys}, 14:\penalty0 473--530,
  1982{\natexlab{b}}.

\bibitem[Hill(1988)]{direct-map-cache}
M.D. Hill.
\newblock A case for direct-mapped caches.
\newblock \emph{Computer}, 21\penalty0 (12), 1988.

\bibitem[Przybylski et~al.(1989)Przybylski, Horowitz, and
  Hennessy]{multi-level-caches}
S.~Przybylski, M.~Horowitz, and J.~Hennessy.
\newblock Characteristics of performance-optimal multi-level cache hierarchies.
\newblock \emph{Proceedings of the 16th Annual International Symposium on
  Computer Architecture}, pages 114--121, 1989.

\bibitem[Jouppi(1990)]{cache-prefetch-buffer}
Norman~P. Jouppi.
\newblock Improving direct-mapped cache performance by the addition of a small
  fully-associative cache and prefetch buffers.
\newblock In \emph{Proceedings of the 17th Annual International Symposium on
  Computer Architecture}, page 364–373, 1990.

\bibitem[Baer and Chen(1991)]{prefetching-1}
Jean-Loup Baer and Tien-Fu Chen.
\newblock An effective on-chip preloading scheme to reduce data access penalty.
\newblock In \emph{Proceedings of the 1991 ACM/IEEE Conference on
  Supercomputing}, pages 176--186, 1991.

\bibitem[Fu et~al.(1992)Fu, Patel, and Janssens]{prefetching-2}
John Fu, Janak Patel, and Bob Janssens.
\newblock Stride directed prefetching in scalar processors.
\newblock \emph{Proceedings the 25th Annual International Symposium on
  Microarchitecture}, pages 102--110, 1992.

\bibitem[Falsafi and Wenisch(2014)]{MC-prefetching}
Babak Falsafi and Thomas~F. Wenisch.
\newblock Morgan and Claypool Publishers, 2014.

\bibitem[Tullsen et~al.(1995)Tullsen, Eggers, and Levy]{smt-1}
Dean~M. Tullsen, Susan~J. Eggers, and Henry~M. Levy.
\newblock Simultaneous multithreading: Maximizing on-chip parallelism.
\newblock In \emph{Proceedings of the 22nd Annual International Symposium on
  Computer Architecture}, page 392–403, 1995.

\bibitem[Tullsen et~al.(1996)Tullsen, Eggers, Emer, Levy, Lo, and Stamm]{smt-2}
Dean~M. Tullsen, Susan~J. Eggers, Joel~S. Emer, Henry~M. Levy, Jack~L. Lo, and
  Rebecca~L. Stamm.
\newblock Exploiting choice: Instruction fetch and issue on an implementable
  simultaneous multithreading processor.
\newblock In \emph{Proceedings of the 23rd Annual International Symposium on
  Computer Architecture}, page 191–202, 1996.

\bibitem[Nemirovsky and Tullsen(2013)]{MC-multithreading}
Mario Nemirovsky and Dean Tullsen.
\newblock Morgan and Claypool Publishers, 2013.

\bibitem[Google({\natexlab{a}})]{datacenter-google-pue}
Google.
\newblock {Google Data Centers Efficiency}.
\newblock \url{https://www.google.com/about/datacenters/efficiency/},
  {\natexlab{a}}.

\bibitem[Facebook()]{datacenter-facebook-pue}
Facebook.
\newblock {Facebook Sustainability -- Data Centers}.
\newblock \url{https://sustainability.fb.com/report-page/data-centers/}.

\bibitem[Lawrence(2019)]{datacenter-pue-uptime-1}
Andy Lawrence.
\newblock {Is PUE actually going UP?}
\newblock \url{https://journal.uptimeinstitute.com/is-pue-actually-going-up/},
  2019.

\bibitem[Lawrence(2020)]{datacenter-pue-uptime}
Andy Lawrence.
\newblock {Data center PUEs flat since 2013}.
\newblock
  \url{https://journal.uptimeinstitute.com/data-center-pues-flat-since-2013/},
  2020.

\bibitem[Wulf and McKee(1996)]{memory-wall}
Wm~Wulf and Sally McKee.
\newblock Hitting the memory wall: Implications of the obvious.
\newblock \emph{Computer Architecture News}, 23, 01 1996.

\bibitem[Dennard et~al.(1974)Dennard, Gaensslen, Yu, Rideout, Bassous, and
  LeBlanc]{power-wall-1}
R.H. Dennard, F.H. Gaensslen, Hwa-Nien Yu, V.L. Rideout, E.~Bassous, and A.R.
  LeBlanc.
\newblock {Design of ion-implanted MOSFET's with very small physical
  dimensions}.
\newblock \emph{IEEE Journal of Solid-State Circuits}, 9\penalty0 (5):\penalty0
  256--268, 1974.

\bibitem[Bohr(2007)]{power-wall-2}
Mark Bohr.
\newblock {A 30 year retrospective on Dennard's MOSFET scaling paper}.
\newblock \emph{IEEE Solid-State Circuits Newsletter}, 12:\penalty0 11--13,
  2007.

\bibitem[Olukotun et~al.(2007)Olukotun, Hammond, and Laudon]{MC-cmp}
Kunle Olukotun, Lance Hammond, and James Laudon.
\newblock Morgan and Claypool Publishers, 2007.

\bibitem[Adve and Gharachorloo(1996)]{consistency-2}
S.V. Adve and K.~Gharachorloo.
\newblock Shared memory consistency models: a tutorial.
\newblock \emph{Computer}, 29\penalty0 (12):\penalty0 66--76, 1996.

\bibitem[Hill(1998)]{consistency-1}
Mark Hill.
\newblock Multiprocessors should support simple memory consistency models.
\newblock \emph{Computer}, 31\penalty0 (8):\penalty0 28--34, 1998.

\bibitem[Sorin et~al.(2011)Sorin, Hill, and Wood]{coherence-consistency}
Daniel~J. Sorin, Mark~D. Hill, and David~A. Wood.
\newblock \emph{A Primer on Memory Consistency and Cache Coherence}.
\newblock Morgan and Claypool Publishers, 2011.
\newblock ISBN 1608455645.

\bibitem[Qureshi et~al.(2007)Qureshi, Jaleel, Patt, Steely, and Emer]{dip}
Moinuddin~K. Qureshi, Aamer Jaleel, Yale~N. Patt, Simon~C. Steely, and Joel
  Emer.
\newblock Adaptive insertion policies for high performance caching.
\newblock In \emph{Proceedings of the 34th Annual International Symposium on
  Computer Architecture}, page 381–391, 2007.

\bibitem[Jaleel et~al.(2010)Jaleel, Theobald, Steely, and Emer]{rrip}
Aamer Jaleel, Kevin~B. Theobald, Simon~C. Steely, and Joel Emer.
\newblock {High Performance Cache Replacement Using Re-Reference Interval
  Prediction (RRIP)}.
\newblock In \emph{Proceedings of the 37th Annual International Symposium on
  Computer Architecture}, page 60–71, 2010.

\bibitem[Wu et~al.(2011)Wu, Jaleel, Hasenplaugh, Martonosi, Steely, and
  Emer]{ship}
Carole-Jean Wu, Aamer Jaleel, Will Hasenplaugh, Margaret Martonosi, Simon~C.
  Steely, and Joel Emer.
\newblock {SHiP: Signature-Based Hit Predictor for High Performance Caching}.
\newblock In \emph{Proceedings of the 44th Annual IEEE/ACM International
  Symposium on Microarchitecture}, page 430–441, 2011.

\bibitem[Balasubramonian et~al.(2011)Balasubramonian, Jouppi, and
  Muralimanohar]{MC-caches}
Rajeev Balasubramonian, Norman~P. Jouppi, and Naveen Muralimanohar.
\newblock Morgan and Claypool Publishers, 2011.

\bibitem[Sardashti et~al.(2015)Sardashti, Arelakis, Stenstom, and
  Wood]{MC-compression}
Somayeh Sardashti, Angelos Arelakis, Per Stenstom, and David~A. Wood.
\newblock Morgan and Claypool Publishers, 2015.

\bibitem[Balasubramonian(2019)]{MC-memory-1}
Rajeev Balasubramonian.
\newblock Morgan and Claypool Publishers, 2019.

\bibitem[Jain and Lin(2019)]{MC-cache-1}
Akanksha Jain and Calvin Lin.
\newblock Morgan and Claypool Publishers, 2019.

\bibitem[Dally and Towles(2001)]{noc-1}
William~J. Dally and Brian Towles.
\newblock Route packets, not wires: On-chip inteconnection networks.
\newblock In \emph{Proceedings of the 38th Annual Design Automation
  Conference}, 2001.

\bibitem[Wang et~al.(2002)Wang, Zhu, Peh, and Malik]{noc-2}
Hang-Sheng Wang, Xinping Zhu, Li-Shiuan Peh, and S.~Malik.
\newblock Orion: a power-performance simulator for interconnection networks.
\newblock In \emph{Proceedings of the 35th Annual IEEE/ACM International
  Symposium on Microarchitecture}, pages 294--305, 2002.

\bibitem[Enright and Peh(2009)]{MC-noc-1}
Natalie Enright and Li-shiuan Peh.
\newblock Morgan and Claypool Publishers, 2009.

\bibitem[Jerger et~al.(2017)Jerger, Krishna, Peh, and Martonosi]{MC-noc-2}
Natalie~Enright Jerger, Tushar Krishna, Li-Shiuan Peh, and Margaret Martonosi.
\newblock Morgan and Claypool Publishers, 2017.

\bibitem[Brooks et~al.(2000)Brooks, Tiwari, and Martonosi]{wattch}
David Brooks, Vivek Tiwari, and Margaret Martonosi.
\newblock Wattch: A framework for architectural-level power analysis and
  optimizations.
\newblock In \emph{Proceedings of the 27th Annual International Symposium on
  Computer Architecture}, pages 83--94, 2000.

\bibitem[Brooks and Martonosi(2001)]{dtm}
D.~Brooks and M.~Martonosi.
\newblock Dynamic thermal management for high-performance microprocessors.
\newblock In \emph{Proceedings of the 7th International Symposium on
  High-Performance Computer Architecture}, pages 171--182, 2001.

\bibitem[Skadron et~al.(2003)Skadron, Stan, Huang, Velusamy, Sankaranarayanan,
  and Tarjan]{temperature-microarchitecture}
K.~Skadron, M.R. Stan, W.~Huang, Sivakumar Velusamy, Karthik Sankaranarayanan,
  and D.~Tarjan.
\newblock Temperature-aware microarchitecture.
\newblock In \emph{Proceedings of the 30th Annual International Symposium on
  Computer Architecture}, pages 2--13, 2003.

\bibitem[Kaxiras and Martonosi(2008)]{MC-power-efficiency}
Stefanos Kaxiras and Margaret Martonosi.
\newblock \emph{Computer Architecture Techniques for Power-Efficiency}.
\newblock Morgan and Claypool Publishers, 1st edition, 2008.
\newblock ISBN 1598292080.

\bibitem[Själander et~al.(2014)Själander, Martonosi, and
  Kaxiras]{MC-power-efficiency-v1}
Magnus Själander, Margaret Martonosi, and Stefanos Kaxiras.
\newblock Morgan and Claypool Publishers, 2014.

\bibitem[Barroso and Hölzle(2009)]{datacenter-computer-1}
Luiz~André Barroso and Urs Hölzle.
\newblock \emph{The Datacenter as a Computer: An Introduction to the Design of
  Warehouse-Scale Machines}.
\newblock 2009.

\bibitem[Greenstein and Fang(2020)]{Greenstein2020WhereTC}
S.~Greenstein and Tommy~Pan Fang.
\newblock Where the cloud rests: The economic geography of data centers.
\newblock 2020.

\bibitem[Esmaeilzadeh et~al.(2011)Esmaeilzadeh, Blem, Amant, Sankaralingam, and
  Burger]{dark-silicon}
Hadi Esmaeilzadeh, Emily Blem, Renée~St. Amant, Karthikeyan Sankaralingam, and
  Doug Burger.
\newblock Dark silicon and the end of multicore scaling.
\newblock In \emph{Proceedings of the 38th Annual International Symposium on
  Computer Architecture}, pages 365--376, 2011.

\bibitem[Jouppi et~al.(2018)Jouppi, Young, Patil, and Patterson]{dsa}
Norman~P. Jouppi, Cliff Young, Nishant Patil, and David Patterson.
\newblock A domain-specific architecture for deep neural networks.
\newblock \emph{{Communication of ACM}}, 61\penalty0 (9):\penalty0 50–59,
  2018.

\bibitem[Ovtcharov et~al.(2015)Ovtcharov, Ruwase, Kim, Fowers, Strauss, and
  Chung]{msft-dnn-accelerator}
Kalin Ovtcharov, Olatunji Ruwase, Joo-Young Kim, Jeremy Fowers, Karin Strauss,
  and Eric Chung.
\newblock Toward accelerating deep learning at scale using specialized hardware
  in the datacenter.
\newblock In \emph{Proceedings of the 27th IEEE HotChips Symposium on
  High-Performance Chips}, 2015.

\bibitem[Chen et~al.(2016)Chen, Emer, and Sze]{Eyeriss}
Yu-Hsin Chen, Joel Emer, and Vivienne Sze.
\newblock Eyeriss: A spatial architecture for energy-efficient dataflow for
  convolutional neural networks.
\newblock In \emph{Proceedings of the ACM/IEEE 43rd Annual International
  Symposium on Computer Architecture}, pages 367--379, 2016.

\bibitem[Jouppi et~al.(2017)Jouppi, Young, Patil, Patterson, Agrawal, Bajwa,
  Bates, Bhatia, Boden, Borchers, Boyle, Cantin, Chao, Clark, Coriell, Daley,
  Dau, Dean, Gelb, Ghaemmaghami, Gottipati, Gulland, Hagmann, Ho, Hogberg, Hu,
  Hundt, Hurt, Ibarz, Jaffey, Jaworski, Kaplan, Khaitan, Killebrew, Koch,
  Kumar, Lacy, Laudon, Law, Le, Leary, Liu, Lucke, Lundin, MacKean, Maggiore,
  Mahony, Miller, Nagarajan, Narayanaswami, Ni, Nix, Norrie, Omernick,
  Penukonda, Phelps, Ross, Ross, Salek, Samadiani, Severn, Sizikov, Snelham,
  Souter, Steinberg, Swing, Tan, Thorson, Tian, Toma, Tuttle, Vasudevan,
  Walter, Wang, Wilcox, and Yoon]{tpu}
Norman~P. Jouppi, Cliff Young, Nishant Patil, David Patterson, Gaurav Agrawal,
  Raminder Bajwa, Sarah Bates, Suresh Bhatia, Nan Boden, Al~Borchers, Rick
  Boyle, Pierre-luc Cantin, Clifford Chao, Chris Clark, Jeremy Coriell, Mike
  Daley, Matt Dau, Jeffrey Dean, Ben Gelb, Tara~Vazir Ghaemmaghami, Rajendra
  Gottipati, William Gulland, Robert Hagmann, C.~Richard Ho, Doug Hogberg, John
  Hu, Robert Hundt, Dan Hurt, Julian Ibarz, Aaron Jaffey, Alek Jaworski,
  Alexander Kaplan, Harshit Khaitan, Daniel Killebrew, Andy Koch, Naveen Kumar,
  Steve Lacy, James Laudon, James Law, Diemthu Le, Chris Leary, Zhuyuan Liu,
  Kyle Lucke, Alan Lundin, Gordon MacKean, Adriana Maggiore, Maire Mahony,
  Kieran Miller, Rahul Nagarajan, Ravi Narayanaswami, Ray Ni, Kathy Nix, Thomas
  Norrie, Mark Omernick, Narayana Penukonda, Andy Phelps, Jonathan Ross, Matt
  Ross, Amir Salek, Emad Samadiani, Chris Severn, Gregory Sizikov, Matthew
  Snelham, Jed Souter, Dan Steinberg, Andy Swing, Mercedes Tan, Gregory
  Thorson, Bo~Tian, Horia Toma, Erick Tuttle, Vijay Vasudevan, Richard Walter,
  Walter Wang, Eric Wilcox, and Doe~Hyun Yoon.
\newblock In-datacenter performance analysis of a tensor processing unit.
\newblock In \emph{Proceedings of the ACM/IEEE 44th Annual International
  Symposium on Computer Architecture}, 2017.

\bibitem[Fowers et~al.(2018)Fowers, Ovtcharov, Papamichael, Massengill, Liu,
  Lo, Alkalay, Haselman, Adams, Ghandi, Heil, Patel, Sapek, Weisz, Woods,
  Lanka, Reinhardt, Caulfield, Chung, and Burger]{catapult}
Jeremy Fowers, Kalin Ovtcharov, Michael Papamichael, Todd Massengill, Ming Liu,
  Daniel Lo, Shlomi Alkalay, Michael Haselman, Logan Adams, Mahdi Ghandi,
  Stephen Heil, Prerak Patel, Adam Sapek, Gabriel Weisz, Lisa Woods, Sitaram
  Lanka, Steve Reinhardt, Adrian Caulfield, Eric Chung, and Doug Burger.
\newblock A configurable cloud-scale dnn processor for real-time ai.
\newblock In \emph{Proceedings of the 45th International Symposium on Computer
  Architecture, 2018}, 2018.

\bibitem[Shao et~al.(2019)Shao, Clemons, Venkatesan, Zimmer, Fojtik, Jiang,
  Keller, Klinefelter, Pinckney, Raina, Tell, Zhang, Dally, Emer, Gray,
  Khailany, and Keckler]{simba}
Yakun~Sophia Shao, Jason Clemons, Rangharajan Venkatesan, Brian Zimmer, Matthew
  Fojtik, Nan Jiang, Ben Keller, Alicia Klinefelter, Nathaniel Pinckney,
  Priyanka Raina, Stephen~G. Tell, Yanqing Zhang, William~J. Dally, Joel Emer,
  C.~Thomas Gray, Brucek Khailany, and Stephen~W. Keckler.
\newblock Simba: Scaling deep-learning inference with multi-chip-module-based
  architecture.
\newblock In \emph{Proceedings of the 52nd Annual IEEE/ACM International
  Symposium on Microarchitecture}, page 14–27, 2019.

\bibitem[Mattson et~al.(2020)Mattson, Reddi, Cheng, Coleman, Diamos, Kanter,
  Micikevicius, Patterson, Schmuelling, Tang, Wei, and Wu]{mlperf-1}
Peter Mattson, Vijay~Janapa Reddi, Christine Cheng, Cody Coleman, Greg Diamos,
  David Kanter, Paulius Micikevicius, David Patterson, Guenther Schmuelling,
  Hanlin Tang, Gu-Yeon Wei, and Carole-Jean Wu.
\newblock Mlperf: An industry standard benchmark suite for machine learning
  performance.
\newblock \emph{IEEE Micro}, 40\penalty0 (2), 2020.

\bibitem[Henry et~al.(2020)Henry, Palangpour, Thomson, Gardner, Arden, Donahue,
  Houck, Johnson, O’Brien, Petersen, Seroussi, and Walker]{DL-accelerators-3}
Glenn Henry, Parviz Palangpour, Michael Thomson, J~Scott Gardner, Bryce Arden,
  Jim Donahue, Kimble Houck, Jonathan Johnson, Kyle O’Brien, Scott Petersen,
  Benjamin Seroussi, and Tyler Walker.
\newblock {High-Performance Deep-Learning Coprocessor Integrated into x86 SoC
  with Server-Class CPUs Industrial Product}.
\newblock In \emph{Proceedings of the ACM/IEEE 47th Annual International
  Symposium on Computer Architecture}, 2020.

\bibitem[Reddi et~al.(2021)Reddi, Cheng, Kanter, Mattson, Schmuelling, and
  Wu]{mlperf-4}
Vijay~Janapa Reddi, Christine Cheng, David Kanter, Peter Mattson, Guenther
  Schmuelling, and Carole-Jean Wu.
\newblock The vision behind mlperf: Understanding ai inference performance.
\newblock \emph{IEEE Micro}, 41\penalty0 (3):\penalty0 10--18, 2021.

\bibitem[Anderson et~al.(2021)Anderson, Chen, Chen, Deng, Fix, Gschwind,
  Kalaiah, Kim, Lee, Liang, Liu, Lu, Montgomery, Moorthy, Nadathur, Naghshineh,
  Nayak, Park, Petersen, Schatz, Sundaram, Tang, Tang, Yang, Yu, Yuen, Zhang,
  Anbudurai, Balan, Bojja, Boyd, Breitbach, Caldato, Calvo, Catron, Chandwani,
  Christeas, Cottel, Coutinho, Dalli, Dhanotia, Duncan, Dzhabarov, Elmir, Fu,
  Fu, Fulthorp, Gangidi, Gibson, Gordon, Hernandez, Ho, Huang, Johansson,
  Juluri, Kanaujia, Kesarkar, Killinger, Kim, Kulkarni, Lele, Li, Li, Li, Liu,
  Liu, Maher, Mallipedi, Mangla, Matam, Mehta, Mehta, Mitchell, Muthiah,
  Nagarkatte, Narasimha, Nguyen, Ortiz, Padmanabha, Pan, Poojary, Ye, Qi,
  Raginel, Rajagopal, Rice, Ross, Rotem, Russ, Shah, Shan, Shen, Shetty,
  Skandakumaran, Srinivasan, Sumbaly, Tauberg, Tzur, Wang, Wang, Wei, Xia, Xu,
  Yang, Zhang, Zhang, Zhao, Zhao, Zhu, Qiao, Smelyanskiy, Jia, and
  Rao]{facebook}
Michael Anderson, Benny Chen, Stephen Chen, Summer Deng, Jordan Fix, Michael
  Gschwind, Aravind Kalaiah, Changkyu Kim, Jaewon Lee, Jason Liang, Haixin Liu,
  Yinghai Lu, Jack Montgomery, Arun Moorthy, Satish Nadathur, Sam Naghshineh,
  Avinash Nayak, Jongsoo Park, Chris Petersen, Martin Schatz, Narayanan
  Sundaram, Bangsheng Tang, Peter Tang, Amy Yang, Jiecao Yu, Hector Yuen, Ying
  Zhang, Aravind Anbudurai, Vandana Balan, Harsha Bojja, Joe Boyd, Matthew
  Breitbach, Claudio Caldato, Anna Calvo, Garret Catron, Sneh Chandwani, Panos
  Christeas, Brad Cottel, Brian Coutinho, Arun Dalli, Abhishek Dhanotia, Oniel
  Duncan, Roman Dzhabarov, Simon Elmir, Chunli Fu, Wenyin Fu, Michael Fulthorp,
  Adi Gangidi, Nick Gibson, Sean Gordon, Beatriz~Padilla Hernandez, Daniel Ho,
  Yu-Cheng Huang, Olof Johansson, Shishir Juluri, Shobhit Kanaujia, Manali
  Kesarkar, Jonathan Killinger, Ben Kim, Rohan Kulkarni, Meghan Lele, Huayu Li,
  Huamin Li, Yueming Li, Cynthia Liu, Jerry Liu, Bert Maher, Chandra Mallipedi,
  Seema Mangla, Kiran~Kumar Matam, Jubin Mehta, Shobhit Mehta, Christopher
  Mitchell, Bharath Muthiah, Nitin Nagarkatte, Ashwin Narasimha, Bernard
  Nguyen, Thiara Ortiz, Soumya Padmanabha, Deng Pan, Ashwin Poojary, Ye, Qi,
  Olivier Raginel, Dwarak Rajagopal, Tristan Rice, Craig Ross, Nadav Rotem,
  Scott Russ, Kushal Shah, Baohua Shan, Hao Shen, Pavan Shetty, Krish
  Skandakumaran, Kutta Srinivasan, Roshan Sumbaly, Michael Tauberg, Mor Tzur,
  Hao Wang, Man Wang, Ben Wei, Alex Xia, Chenyu Xu, Martin Yang, Kai Zhang,
  Ruoxi Zhang, Ming Zhao, Whitney Zhao, Rui Zhu, Lin Qiao, Misha Smelyanskiy,
  Bill Jia, and Vijay Rao.
\newblock First-generation inference accelerator deployment at {Facebook},
  2021.

\bibitem[Jang et~al.(2021)Jang, Lee, Kim, Park, Ardestani, Choi, Kim, Kim, Yu,
  Abdel-Aziz, Park, Lee, Dongwoo, Lee, Kim, Jung, Nam, Lim, Lee, Song, Kwon,
  Hassoun, Lim, and Choi]{DL-accelerators-1}
Jun-Woo Jang, Sehwan Lee, Dongyoung Kim, Hyunsung Park, Ali~Shafiee Ardestani,
  Yeongjae Choi, Channoh Kim, Yoojin Kim, Hyeongseok Yu, Hamzah Abdel-Aziz,
  Jun-Seok Park, H.~Lee, Dongwoo, Lee, Myeong~Woo Kim, Hanwoong Jung, Hee-Young
  Nam, Dong-Hyuk Lim, Seungwon Lee, Joonho Song, Suk-Chon Kwon, Joseph Hassoun,
  SukHwan Lim, and Changkyu Choi.
\newblock Sparsity-aware and re-configurable npu architecture for samsung
  flagship mobile soc.
\newblock In \emph{Proceedings of the ACM/IEEE 48th Annual International
  Symposium on Computer Architecture}, 2021.

\bibitem[Thompto et~al.(2021)Thompto, Nguyen, Moreira, Bertran, Jacobson,
  Eickemeyer, Rao, Goulet, Byers, Gonzalez, Swaminathan, Dhanwada, M{\"u}ller,
  Wagner, Sadasivam, Montoye, Starke, Zoellin, Floyd, Stuecheli,
  Chandramoorthy, Wellman, Buyuktosunoglu, Pflanz, Sinharoy, and
  Bose]{DL-accelerators-2}
Brian~W. Thompto, Dung~Q. Nguyen, J.~Moreira, Ramon Bertran, H.~Jacobson,
  R.~Eickemeyer, R.~Rao, M.~Goulet, Marcy Byers, Christopher~J. Gonzalez,
  Karthik Swaminathan, N.~Dhanwada, Silvia~M. M{\"u}ller, Andreas Wagner,
  S.~Sadasivam, R.~Montoye, William~J. Starke, Christian~G. Zoellin, M.~Floyd,
  Jeffrey Stuecheli, N.~Chandramoorthy, J.~Wellman, A.~Buyuktosunoglu,
  M.~Pflanz, B.~Sinharoy, and P.~Bose.
\newblock Energy efficiency boost in the ai-infused power 10 processor:
  Industrial product.
\newblock In \emph{Proceedings of the ACM/IEEE 48th Annual International
  Symposium on Computer Architecture}, 2021.

\bibitem[{NVIDIA}()]{gpu}
{NVIDIA}.
\newblock {Tensor Cores: Unprecedented Acceleration for HPC and AI}.
\newblock \url{https://www.nvidia.com/en-us/data-center/tensor-cores/}.

\bibitem[Taylor et~al.(2020)Taylor, Vega, Khazraee, Magaki, Davidson, and
  Richmond]{cloud-accelerators-1}
Michael Taylor, Luis Vega, Moein Khazraee, Ikuo Magaki, Scott Davidson, and
  Dustin Richmond.
\newblock Asic clouds: specializing the datacenter for planet-scale
  applications.
\newblock \emph{Communications of the ACM}, 63:\penalty0 103--109, 06 2020.
\newblock \doi{10.1145/3399734}.

\bibitem[Ranganathan et~al.(2021)Ranganathan, Stodolsky, Calow, Dorfman,
  Guevara, Smullen~IV, Kuusela, Balasubramanian, Bhatia, Chauhan, Cheung,
  Chong, Dasharathi, Feng, Fosco, Foss, Gelb, Gwin, Hase, He, Ho, Huffman~Jr.,
  Indupalli, Jayaram, Kongetira, Kyaw, Laursen, Li, Lou, Lucke, Maaninen,
  Macias, Mahony, Munday, Muroor, Penukonda, Perkins-Argueta, Persaud, Ramirez,
  Rautio, Ripley, Salek, Sekar, Sokolov, Springer, Stark, Tan, Wachsler,
  Walton, Wickeraad, Wijaya, and Wu]{cloud-accelerators-2}
Parthasarathy Ranganathan, Daniel Stodolsky, Jeff Calow, Jeremy Dorfman,
  Marisabel Guevara, Clinton~Wills Smullen~IV, Aki Kuusela, Raghu
  Balasubramanian, Sandeep Bhatia, Prakash Chauhan, Anna Cheung, In~Suk Chong,
  Niranjani Dasharathi, Jia Feng, Brian Fosco, Samuel Foss, Ben Gelb, Sara~J.
  Gwin, Yoshiaki Hase, Da-ke He, C.~Richard Ho, Roy~W. Huffman~Jr., Elisha
  Indupalli, Indira Jayaram, Poonacha Kongetira, Cho~Mon Kyaw, Aaron Laursen,
  Yuan Li, Fong Lou, Kyle~A. Lucke, JP~Maaninen, Ramon Macias, Maire Mahony,
  David~Alexander Munday, Srikanth Muroor, Narayana Penukonda, Eric
  Perkins-Argueta, Devin Persaud, Alex Ramirez, Ville-Mikko Rautio, Yolanda
  Ripley, Amir Salek, Sathish Sekar, Sergey~N. Sokolov, Rob Springer, Don
  Stark, Mercedes Tan, Mark~S. Wachsler, Andrew~C. Walton, David~A. Wickeraad,
  Alvin Wijaya, and Hon~Kwan Wu.
\newblock Warehouse-scale video acceleration: Co-design and deployment in the
  wild.
\newblock In \emph{Proceedings of the 26th ACM International Conference on
  Architectural Support for Programming Languages and Operating Systems}, page
  600–615, 2021.

\bibitem[Hamari et~al.(2016)Hamari, Sjöklint, and Ukkonen]{sharing-economy}
Juho Hamari, Mimmi Sjöklint, and Antti Ukkonen.
\newblock The sharing economy: Why people participate in collaborative
  consumption.
\newblock \emph{Journal of the Association for Information Science and
  Technology}, 67:\penalty0 2047--2059, 09 2016.

\bibitem[Sun et~al.(2019)Sun, Agostini, Dong, and Kaeli]{cpu-gpu-ppw}
Yifan Sun, Nicolas~Bohm Agostini, Shi Dong, and D.~Kaeli.
\newblock Summarizing cpu and gpu design trends with product data.
\newblock \emph{ArXiv}, abs/1911.11313, 2019.

\bibitem[Simons and Jones(2012)]{societal-challenge-10}
Barbara Simons and Douglas~W. Jones.
\newblock Internet voting in the u.s.
\newblock \emph{Communication of ACM}, 55\penalty0 (10):\penalty0 68–77,
  2012.

\bibitem[Abelson et~al.(2015)Abelson, Anderson, Bellovin, Benaloh, Blaze,
  Diffie, Gilmore, Green, Landau, Neumann, Rivest, Schiller, Schneier, Specter,
  and Weitzner]{societal-challenge-11}
H.~Abelson, Ross~J. Anderson, S.~Bellovin, Josh Benaloh, M.~Blaze, W.~Diffie,
  J.~Gilmore, Matthew Green, S.~Landau, P.~Neumann, R.~Rivest, J.~Schiller,
  B.~Schneier, Michael~A. Specter, and D.~Weitzner.
\newblock Keys under doormats: mandating insecurity by requiring government
  access to all data and communications.
\newblock \emph{Journal of Cybersecurity}, 1:\penalty0 69--79, 2015.

\bibitem[Gervais et~al.(2016)Gervais, Karame, W\"{u}st, Glykantzis, Ritzdorf,
  and Capkun]{societal-challenge-12}
Arthur Gervais, Ghassan~O. Karame, Karl W\"{u}st, Vasileios Glykantzis, Hubert
  Ritzdorf, and Srdjan Capkun.
\newblock On the security and performance of proof of work blockchains.
\newblock In \emph{Proceedings of the ACM Conference on Computer and
  Communications Security}, page 3–16, 2016.

\bibitem[Ahmed et~al.(2017)Ahmed, Haque, Chen, and Dell]{societal-challenge-13}
Syed~Ishtiaque Ahmed, Md.~Romael Haque, Jay Chen, and Nicola Dell.
\newblock Digital privacy challenges with shared mobile phone use in
  bangladesh.
\newblock \emph{Proceedings of the ACM Human-Computer Interaction}, 1\penalty0
  (CSCW), 2017.

\bibitem[Whittaker et~al.(2018)Whittaker, Crawford, Dobbe, Fried, Kaziunas,
  Mathur, West, Richardson, Schultz, and Schwartz]{ai-now-2018}
Meredith Whittaker, Kate Crawford, Roel Dobbe, Genevieve Fried, Elizabeth
  Kaziunas, Varoon Mathur, Sarah~Mysers West, Rashida Richardson, Jason
  Schultz, and Oscar Schwartz.
\newblock {AI Now 2018 Report}.
\newblock
  \url{https://ec.europa.eu/futurium/en/system/files/ged/ai_now_2018_report.pdf},
  2018.

\bibitem[Speicher et~al.(2018)Speicher, Ali, Venkatadri, Ribeiro, Arvanitakis,
  Benevenuto, Gummadi, Loiseau, and Mislove]{societal-challenge-8}
Till Speicher, Muhammad Ali, Giridhari Venkatadri, Filipe~Nunes Ribeiro, George
  Arvanitakis, Fabrício Benevenuto, Krishna~P. Gummadi, Patrick Loiseau, and
  Alan Mislove.
\newblock Potential for discrimination in online targeted advertising.
\newblock In \emph{Proceedings of the Conference on Fairness, Accountability
  and Transparency}, pages 5--19, 2018.

\bibitem[Chouldechova et~al.(2018)Chouldechova, Benavides-Prado, Fialko, and
  Vaithianathan]{societal-challenge-7}
Alexandra Chouldechova, Diana Benavides-Prado, Oleksandr Fialko, and Rhema
  Vaithianathan.
\newblock A case study of algorithm-assisted decision making in child
  maltreatment hotline screening decisions.
\newblock In \emph{Proceedings of the Conference on Fairness, Accountability
  and Transparency}, pages 134--148, 2018.

\bibitem[Ekstrand et~al.(2018)Ekstrand, Tian, Azpiazu, Ekstrand, Anuyah,
  McNeill, and Pera]{societal-challenge-6}
Michael Ekstrand, Mucun Tian, Ion Azpiazu, Jennifer Ekstrand, Oghenemaro
  Anuyah, David McNeill, and Maria Pera.
\newblock All the cool kids, how do they fit in?: Popularity and demographic
  biases in recommender evaluation and effectiveness.
\newblock In \emph{Proceedings of the Conference on Fairness, Accountability,
  and Transparency}, 2018.

\bibitem[Selbst et~al.(2019)Selbst, Boyd, Friedler, Venkatasubramanian, and
  Vertesi]{societal-challenge-4}
Andrew~D. Selbst, Danah Boyd, Sorelle~A. Friedler, Suresh Venkatasubramanian,
  and Janet Vertesi.
\newblock Fairness and abstraction in sociotechnical systems.
\newblock In \emph{Proceedings of the Conference on Fairness, Accountability,
  and Transparency}, 2019.

\bibitem[Ali et~al.(2019)Ali, Sapiezynski, Bogen, Korolova, Mislove, and
  Rieke]{societal-challenge-8a}
Muhammad Ali, Piotr Sapiezynski, Miranda Bogen, Aleksandra Korolova, Alan
  Mislove, and Aaron Rieke.
\newblock Discrimination through optimization: How facebook's ad delivery can
  lead to biased outcomes.
\newblock 2019.

\bibitem[Babaei et~al.(2019)Babaei, Chakraborty, Kulshrestha, Redmiles, Cha,
  and Gummadi]{societal-challenge-5}
Mahmoudreza Babaei, Abhijnan Chakraborty, Juhi Kulshrestha, Elissa~M. Redmiles,
  Meeyoung Cha, and Krishna~P. Gummadi.
\newblock Analyzing biases in perception of truth in news stories and their
  implications for fact checking.
\newblock In \emph{Proceedings of the Conference on Fairness, Accountability,
  and Transparency}, 2019.

\bibitem[Crawford et~al.(2019)Crawford, Dobbe, Dryer, Fried, Green, Kaziunas,
  Kak, Mathur, McElroy, Sánchez, Raji, Rankin, Richardson, Schultz, West, and
  Whittaker]{ai-now-2019}
Kate Crawford, Roel Dobbe, Theodora Dryer, Genevieve Fried, Ben Green,
  Elizabeth Kaziunas, Amba Kak, Varoon Mathur, Erin McElroy, Andrea~Nill
  Sánchez, Deborah Raji, Joy~Lisi Rankin, Rashida Richardson, Jason Schultz,
  Sarah~Myers West, and Meredith Whittaker.
\newblock {AI Now 2019 Report}.
\newblock \url{https://ainowinstitute.org/AI_Now_2019_Report.pdf}, 2019.

\bibitem[Papakyriakopoulos et~al.(2020)Papakyriakopoulos, Hegelich, Serrano,
  and Marco]{societal-challenge-2}
Orestis Papakyriakopoulos, Simon Hegelich, Juan Carlos~Medina Serrano, and
  Fabienne Marco.
\newblock Bias in word embeddings.
\newblock In \emph{Proceedings of the 2020 Conference on Fairness,
  Accountability, and Transparency}, page 446–457, 2020.

\bibitem[Park et~al.(2021)Park, Specter, Narula, and
  Rivest]{societal-challenge-9}
Sunoo Park, Michael~A. Specter, Neha Narula, and R.~Rivest.
\newblock Going from bad to worse: from internet voting to blockchain voting.
\newblock \emph{Journal of Cybersecurity}, 7, 2021.

\bibitem[Bender et~al.(2021)Bender, Gebru, McMillan-Major, and
  Shmitchell]{societal-challenge-1}
Emily~M. Bender, Timnit Gebru, Angelina McMillan-Major, and Shmargaret
  Shmitchell.
\newblock On the dangers of stochastic parrots: Can language models be too big?
\newblock In \emph{Proceedings of the 2021 ACM Conference on Fairness,
  Accountability, and Transparency}, page 610–623, 2021.

\bibitem[Kleinberg and Raghavan(2021)]{societal-challenge-3}
Jon Kleinberg and Manish Raghavan.
\newblock Algorithmic monoculture and social welfare.
\newblock \emph{Proceedings of the National Academy of Sciences}, 118\penalty0
  (22), 2021.

\bibitem[Jain and Wullert(2002)]{ict-footprint-5}
Ravi Jain and John Wullert.
\newblock Challenges: Environmental design for pervasive computing systems.
\newblock In \emph{Proceedings of the 8th Annual International Conference on
  Mobile Computing and Networking}, page 263–270, 2002.

\bibitem[Strubell et~al.(2019)Strubell, Ganesh, and McCallum]{ict-footprint}
Emma Strubell, Ananya Ganesh, and Andrew McCallum.
\newblock Energy and policy considerations for deep learning in nlp, 2019.

\bibitem[Manne(2020)]{ict-footprint-2}
Srilatha Manne.
\newblock {Examining the Carbon Footprint of Devices}.
\newblock
  \url{https://devblogs.microsoft.com/sustainable-software/examining-the-carbon-footprint-of-devices/},
  2020.

\bibitem[Wu and Gupta(2021)]{ict-footprint-3}
Carole-Jean Wu and Udit Gupta.
\newblock {Most of computing’s carbon emissions are coming from manufacturing
  and infrastructure}.
\newblock \url{https://tech.fb.com/sustainable-computing/}, 2021.

\bibitem[Gupta et~al.(2021)Gupta, Kim, Lee, Tse, Lee, Wei, Brooks, and
  Wu]{ict-footprint-4}
Udit Gupta, Young~Geun Kim, S.~Lee, J.~Tse, Hsien-Hsin~S. Lee, Gu-Yeon Wei,
  D.~Brooks, and Carole-Jean Wu.
\newblock Chasing carbon: The elusive environmental footprint of computing.
\newblock \emph{Proceedings of the IEEE International Symposium on
  High-Performance Computer Architecture}, pages 854--867, 2021.

\bibitem[Crawford et~al.(2021)Crawford, King, and Wu]{ict-footprint-1}
Alan Crawford, Ian King, and Debby Wu.
\newblock {The Chip Industry Has a Problem With Its Giant Carbon Footprint}.
\newblock
  \url{https://www.bloomberg.com/news/articles/2021-04-08/the-chip-industry-has-a-problem-with-its-giant-carbon-footprint},
  2021.

\bibitem[Lin(2021)]{internet-statistics}
Ying Lin.
\newblock {10 Internet Statistics Every Marketer Should Know in 2021}.
\newblock \url{https://www.oberlo.com/blog/internet-statistics}, May 2021.

\bibitem[Wheeler(2020)]{fcc-data}
Tom Wheeler.
\newblock {5 steps to get the internet to all americans}.
\newblock
  \url{https://www.brookings.edu/research/5-steps-to-get-the-internet-to-all-americans/},
  May 2020.

\bibitem[unicef(2020)]{unicef}
unicef.
\newblock {Two thirds of the world’s school-age children have no internet
  access at home.}
\newblock
  \url{https://www.unicef.org/press-releases/two-thirds-worlds-school-age-children-have-no-internet-access-home-new-unicef-itu},
  2020.

\bibitem[Microsoft({\natexlab{a}})]{microsoft-airband}
Microsoft.
\newblock {Airband: The initiative to bring the internet to everyone}.
\newblock
  \url{https://news.microsoft.com/on-the-issues/2020/09/01/airband-initiative-rural-broadband-digital-divide/},
  {\natexlab{a}}.

\bibitem[Loon()]{google-loon}
Loon.
\newblock {Loon: Expanding internet connectivity with stratospheric balloons}.
\newblock \url{https://x.company/projects/loon/}.

\bibitem[Wikipedia()]{facebook-aquila}
Wikipedia.
\newblock {Facebook Aquila}.
\newblock \url{https://en.wikipedia.org/wiki/Facebook_Aquila}.

\bibitem[{Starlink}()]{starlink}
{Starlink}.
\newblock {High-speed, low latency broadband internet}.
\newblock \url{https://www.starlink.com/}.

\bibitem[Lipp et~al.(2018)Lipp, Schwarz, Gruss, Prescher, Haas, Fogh, Horn,
  Mangard, Kocher, Genkin, Yarom, and Hamburg]{meltdown}
Moritz Lipp, Michael Schwarz, Daniel Gruss, Thomas Prescher, Werner Haas,
  Anders Fogh, Jann Horn, Stefan Mangard, Paul Kocher, Daniel Genkin, Yuval
  Yarom, and Mike Hamburg.
\newblock Meltdown: Reading kernel memory from user space.
\newblock In \emph{Proceedings of the 27th {USENIX} Security Symposium}, 2018.

\bibitem[Kocher et~al.(2019)Kocher, Horn, Fogh, , Genkin, Gruss, Haas, Hamburg,
  Lipp, Mangard, Prescher, Schwarz, and Yarom]{spectre}
Paul Kocher, Jann Horn, Anders Fogh, , Daniel Genkin, Daniel Gruss, Werner
  Haas, Mike Hamburg, Moritz Lipp, Stefan Mangard, Thomas Prescher, Michael
  Schwarz, and Yuval Yarom.
\newblock Spectre attacks: Exploiting speculative execution.
\newblock In \emph{Proceedings of the 40th IEEE Symposium on Security and
  Privacy}, 2019.

\bibitem[Microsoft(2021{\natexlab{a}})]{Msft-redundant-storage}
Microsoft.
\newblock {Azure Storage redundancy}.
\newblock
  \url{https://docs.microsoft.com/en-us/azure/storage/common/storage-redundancy},
  July 2021{\natexlab{a}}.

\bibitem[Microsoft(2021{\natexlab{b}})]{Msft-redundant-network}
Microsoft.
\newblock {Designing for disaster recovery with ExpressRoute private peering}.
\newblock
  \url{https://docs.microsoft.com/en-us/azure/expressroute/designing-for-disaster-recovery-with-expressroute-privatepeering},
  March 2021{\natexlab{b}}.

\bibitem[Sverdlik(2021)]{texasevent}
Yevegeniy Sverdlik.
\newblock {Most Texas Data Centers Weathered the Storm, But Things Did Not Go
  Smoothly}.
\newblock
  \url{https://www.datacenterknowledge.com/uptime/most-texas-data-centers-weathered-storm-things-did-not-go-smoothly},
  March 2021.

\bibitem[{Department of Energy}(2019)]{doe-microgrid}
{Department of Energy}.
\newblock {Designing And Managing Data Centers For Resilience: Demand Response
  And Microgrids}.
\newblock
  \url{https://www.wbdg.org/continuing-education/femp-courses/fempodw034},
  2019.

\bibitem[Roth(2021)]{oregon-fire}
Sammy Roth.
\newblock {How an Oregon wildfire almost derailed California’s power grid}.
\newblock
  \url{https://docs.microsoft.com/en-us/azure/storage/common/storage-redundancy},
  July 2021.

\bibitem[Hornsby(2018)]{resilient-1}
Adrian Hornsby.
\newblock Patterns for resilient architecture — part 1: The story of
  embracing failure at scale.
\newblock
  \url{https://medium.com/the-cloud-architect/patterns-for-resilient-architecture-part-1-d3b60cd8d2b6},
  2018.

\bibitem[{Google Cloud Architecture Center}()]{resilient-2}
{Google Cloud Architecture Center}.
\newblock {Patterns for scalable and resilient apps}.
\newblock
  \url{https://cloud.google.com/architecture/scalable-and-resilient-apps}.

\bibitem[{The World Bank}(2020)]{africa-energy}
{The World Bank}.
\newblock {Lighting Up Africa: Bringing Renewable, Off-Grid Energy to
  Communities}.
\newblock
  \url{https://www.worldbank.org/en/news/feature/2020/08/13/lighting-up-africa-bringing-renewable-off-grid-energy-to-communities},
  August 2020.

\bibitem[Íñigo Goiri et~al.(2015)Íñigo Goiri, Haque, Le, Beauchea, Nguyen,
  Guitart, Torres, and Bianchini]{GreenSlot}
Íñigo Goiri, Md~E. Haque, Kien Le, Ryan Beauchea, Thu~D. Nguyen, Jordi
  Guitart, Jordi Torres, and Ricardo Bianchini.
\newblock Matching renewable energy supply and demand in green datacenters.
\newblock \emph{Ad Hoc Networks}, 25:\penalty0 520--534, 2015.

\bibitem[Radovanovic et~al.(2021)Radovanovic, Koningstein, Schneider, Chen,
  Duarte, Roy, Xiao, Haridasan, Hung, Care, Talukdar, Mullen, Smith, Cottman,
  and Cirne]{carbon-aware-computing}
Ana Radovanovic, Ross Koningstein, Ian Schneider, Bokan Chen, Alexandre Duarte,
  Binz Roy, Diyue Xiao, Maya Haridasan, Patrick Hung, Nick Care, Saurav
  Talukdar, Eric Mullen, Kendal Smith, MariEllen Cottman, and Walfredo Cirne.
\newblock Carbon-aware computing for datacenters.
\newblock \emph{CoRR}, abs/2106.11750, 2021.
\newblock URL \url{https://arxiv.org/abs/2106.11750}.

\bibitem[Lin et~al.(2021)Lin, Zavala, and Chien]{dc-grid-coupling}
Liuzixuan Lin, Victor~M. Zavala, and Andrew~A. Chien.
\newblock Evaluating coupling models for cloud datacenters and power grids.
\newblock In \emph{Proceedings of the 12th ACM International Conference on
  Future Energy Systems}, 2021.

\bibitem[Zhang et~al.(2021)Zhang, Kumbhare, Manousakis, Zhang, Misra, Assis,
  Woolcock, Mahalingam, Warrier, Gauthier, Kunnath, Solomon, Morales, Fontoura,
  and Bianchini]{Flex}
Chaojie Zhang, Alok Kumbhare, Ioannis Manousakis, Deli Zhang, Pulkit Misra, Rod
  Assis, Kyle Woolcock, Nithish Mahalingam, Brijesh Warrier, David Gauthier,
  Lalu Kunnath, Steve Solomon, Osvaldo Morales, Marcus Fontoura, and Ricardo
  Bianchini.
\newblock Flex: High-availability datacenters with zero reserved power.
\newblock In \emph{Proceedings of the International Symposium on Computer
  Architecture}, 2021.

\bibitem[{Facebook}()]{facebook-sustainability}
{Facebook}.
\newblock {Facebook is committed to reaching net zero emissions across our
  value chain in 2030, aligning our efforts with the latest science on what is
  needed to transition to a zero-carbon future}.
\newblock \url{https://sustainability.fb.com/report-page/climate/}.

\bibitem[Apple()]{apple-sustainability}
Apple.
\newblock {Apple commits to be 100 percent carbon neutral for its supply chain
  and products by 2030}.
\newblock
  \url{https://www.apple.com/newsroom/2020/07/apple-commits-to-be-100-percent-carbon-neutral-for-its-supply-chain-and-products-by-2030/}.

\bibitem[Smith()]{msft-sustainability}
Brad Smith.
\newblock {Microsoft will be carbon negative by 2030}.
\newblock
  \url{https://blogs.microsoft.com/blog/2020/01/16/microsoft-will-be-carbon-negative-by-2030/}.

\bibitem[Google({\natexlab{b}})]{google-sustainability}
Google.
\newblock {The Internet is 24x7—carbon-free energy should be too}.
\newblock \url{https://sustainability.google/progress/projects/24x7/},
  {\natexlab{b}}.

\bibitem[Schechner(2021)]{renewable-purchase}
Sam Schechner.
\newblock {Amazon and Other Tech Giants Race to Buy Up Renewable Energy}.
\newblock
  \url{https://www.wsj.com/articles/amazon-and-other-tech-giants-race-to-buy-up-renewable-energy-11624438894},
  2021.

\bibitem[Myers(2020)]{china-climate}
Steven~Lee Myers.
\newblock {China’s Pledge to Be Carbon Neutral by 2060: What It Means}.
\newblock
  \url{https://www.nytimes.com/2020/09/23/world/asia/china-climate-change.html},
  September 2020.

\bibitem[{BBC}(2021)]{eu-climate}
{BBC}.
\newblock {Climate change: EU to cut CO2 emissions by 55\% by 2030}.
\newblock \url{https://www.bbc.com/news/world-europe-56828383}, April 2021.

\bibitem[Orcuttarchive(2015)]{bio-degradable}
Mike Orcuttarchive.
\newblock A biodegradable computer chip that performs surprisingly well.
\newblock
  \url{https://www.technologyreview.com/2015/07/14/167161/a-biodegradable-computer-chip-that-performs-surprisingly-well/},
  2015.

\bibitem[Chang et~al.(2017)Chang, Yao, Jackson, Rand, and
  Wentzlaff]{bio-degradable-micro}
Ting-Jung Chang, Zhuozhi Yao, Paul~J. Jackson, Barry~P. Rand, and David
  Wentzlaff.
\newblock Architectural tradeoffs for biodegradable computing.
\newblock In \emph{Proceedings of the 50th Annual IEEE/ACM International
  Symposium on Microarchitecture}, page 706–717, 2017.

\bibitem[Amazon(2019)]{aws-cloud}
Amazon.
\newblock {The Carbon Reduction Opportunity of Moving to Amazon Web Services}.
\newblock
  \url{https://sustainability.aboutamazon.com/carbon_reduction_aws.pdf}, 2019.

\bibitem[Microsoft(2020)]{msft-cloud}
Microsoft.
\newblock {The Carbon Benefits of Cloud Computing: a Study of the Microsoft
  Cloud}.
\newblock \url{https://www.microsoft.com/en-us/download/details.aspx?id=56950},
  2020.

\bibitem[Evans and Gao(2016)]{google-cloud}
Richard Evans and Jim Gao.
\newblock {DeepMind AI Reduces Google Data Centre Cooling Bill by 40\%}.
\newblock
  \url{https://deepmind.com/blog/article/deepmind-ai-reduces-google-data-centre-cooling-bill-40},
  2016.

\bibitem[Lee and Rowe(2020)]{facebook-cloud}
Dan Lee and Jonathan Rowe.
\newblock {Software, servers, systems, sensors, and science: Facebook’s
  recipe for hyperefficient data centers}.
\newblock \url{https://tech.fb.com/hyperefficient-data-centers/}, 2020.

\bibitem[Masanet et~al.(2020)Masanet, Shehabi, Lei, Smith, and Koomey]{Masanet}
Eric Masanet, Arman Shehabi, Nuoa Lei, Sarah Smith, and Jonathan Koomey.
\newblock Recalibrating global data center energy-use estimates.
\newblock \emph{Science}, 367\penalty0 (6481):\penalty0 984--986, 2020.

\bibitem[Cordella et~al.(2020)Cordella, Alfieri, Clemm, and
  Berwald]{phonedurability}
Mauro Cordella, Felice Alfieri, Christian Clemm, and Anton Berwald.
\newblock Durability of smartphones: A technical analysis of reliability and
  repairability aspects.
\newblock \emph{Journal of Cleaner Production}, 286:\penalty0 125388, 12 2020.
\newblock \doi{10.1016/j.jclepro.2020.125388}.

\bibitem[Ascierto and Lawrence(2020)]{uptimesurvey}
Rhonda Ascierto and Andy Lawrence.
\newblock {Uptime Institute global data center survey 2020}.
\newblock
  \url{https://uptimeinstitute.com/2020-data-center-industry-survey-results},
  2020.

\bibitem[{Fairphone}()]{fair-phone}
{Fairphone}.
\newblock {The world's most sustainable smartphone now at a lower price}.
\newblock \url{https://www.fairphone.com/en/}.

\bibitem[Wiens and Gordon-Byrne(2017)]{right-to-repair}
Kyle Wiens and Gay Gordon-Byrne.
\newblock {Why We Must Fight for the Right to Repair Our Electronics}.
\newblock
  \url{https://spectrum.ieee.org/green-tech/conservation/why-we-must-fight-for-the-right-to-repair-our-electronics},
  2017.

\bibitem[Alsever(2021)]{exec-order}
Jennifer Alsever.
\newblock {What Biden's 'right-to-repair' order could mean for Apple and
  Tesla}.
\newblock
  \url{https://fortune.com/2021/07/09/right-to-repair-order-biden-apple-tesla-hacks/},
  July 2021.

\bibitem[Microsoft(2015)]{farm-beats}
Microsoft.
\newblock {FarmBeats: Democratizing AI for farmers around the world}.
\newblock \url{https://www.microsoft.com/en-us/garage/wall-of-fame/farmbeats/},
  2015.

\bibitem[Gates(2021)]{gates}
Bill Gates.
\newblock \emph{{How to Avoid a Climate Disaster: The Solutions We Have and the
  Breakthroughs We Need}}.
\newblock Knopf, 2021.

\bibitem[{U.S. Energy Information Adminstration}()]{us-electricity-use}
{U.S. Energy Information Adminstration}.
\newblock {Electricity explained: Use of electricity}.
\newblock
  \url{https://www.eia.gov/energyexplained/electricity/use-of-electricity.php}.

\bibitem[Zitnick et~al.(2020)Zitnick, Chanussot, Das, Goyal, Heras-Domingo, Ho,
  Hu, Lavril, Palizhati, Riviere, Shuaibi, Sriram, Tran, Wood, Yoon, Parikh,
  and Ulissi]{open-catalyst}
C.~Lawrence Zitnick, Lowik Chanussot, Abhishek Das, Siddharth Goyal, Javier
  Heras-Domingo, Caleb Ho, Weihua Hu, Thibaut Lavril, Aini Palizhati, Morgane
  Riviere, Muhammed Shuaibi, Anuroop Sriram, Kevin Tran, Brandon Wood, Junwoong
  Yoon, Devi Parikh, and Zachary Ulissi.
\newblock An introduction to electrocatalyst design using machine learning for
  renewable energy storage, 2020.

\bibitem[Elkin and Witherspoon()]{AI-load-shaping}
Carl Elkin and Sims Witherspoon.
\newblock {Machine learning can boost the value of wind energy}.
\newblock
  \url{https://deepmind.com/blog/article/machine-learning-can-boost-value-wind-energy}.

\bibitem[Tollefson(2021)]{covid-co2}
Jeff Tollefson.
\newblock {COVID curbed carbon emissions in 2020 — but not by much}.
\newblock \url{https://www.nature.com/articles/d41586-021-00090-3}, 2021.

\bibitem[Chang et~al.(2010)Chang, Meza, Ranganathan, Bash, and
  Shah]{green-server-design}
Jichuan Chang, Justin Meza, P.~Ranganathan, C.~Bash, and Amip Shah.
\newblock Green server design: beyond operational energy to sustainability.
\newblock In \emph{Proceedings of the 2010 international conference on Power
  aware computing and systems}, 2010.

\bibitem[Bardon et~al.(2020)Bardon, Wuytens, Ragnarsson, Mirabelli, Jang,
  Willems, Mallik, Spessot, Ryckaert, and Parvais]{imec-iedm}
M.~Bardon, P.~Wuytens, L.-Å. Ragnarsson, G.~Mirabelli, D.~Jang, G.~Willems,
  A.~Mallik, A.~Spessot, J.~Ryckaert, and B.~Parvais.
\newblock Dtco including sustainability:
  Power-performance-area-cost-environmental score (ppace) analysis for logic
  technologies.
\newblock In \emph{Proceedings of the IEEE International Electron Devices
  Meeting}, 2020.

\bibitem[Patterson et~al.(2021)Patterson, Gonzalez, Le, Liang, Munguia,
  Rothchild, So, Texier, and Dean]{ai-footprint}
David Patterson, Joseph Gonzalez, Quoc Le, Chen Liang, Lluis-Miquel Munguia,
  Daniel Rothchild, David So, Maud Texier, and Jeff Dean.
\newblock Carbon emissions and large neural network training, 2021.

\bibitem[Hager et~al.(2019)Hager, Drobnis, Fang, Ghani, Greenwald, Lyons,
  Parkes, Schultz, Saria, Smith, and Tambe]{ai-social-good-2}
Gregory~D. Hager, Ann Drobnis, Fei Fang, Rayid Ghani, Amy Greenwald, Terah
  Lyons, David~C. Parkes, Jason Schultz, Suchi Saria, Stephen~F. Smith, and
  Milind Tambe.
\newblock Artificial intelligence for social good, 2019.

\bibitem[Tomasev et~al.(2020)Tomasev, Cornebise, Hutter, Mohamed, Picciariello,
  Connelly, Belgrave, Ezer, van~der Haert, Mugisha, Abila, Arai, Almiraat,
  Proskurnia, Snyder, Otake-Matsuura, Othman, Glasmachers, Wever, Teh, Khan,
  Winne, Schaul, and Clopath]{ai-social-good-1}
Nenad Tomasev, Julien Cornebise, F.~Hutter, S.~Mohamed, Angela Picciariello,
  Bec Connelly, D.~Belgrave, Daphne Ezer, Fanny~Cachat van~der Haert, Frank
  Mugisha, G.~Abila, Hiromi Arai, Hisham Almiraat, Julia Proskurnia, Kyle
  Snyder, M.~Otake-Matsuura, M.~Othman, T.~Glasmachers, W.~D. Wever, Y.~Teh,
  M.~E. Khan, Ruben~De Winne, Tom Schaul, and C.~Clopath.
\newblock {AI} for social good: unlocking the opportunity for positive impact.
\newblock \emph{Nature Communications}, 11, 2020.

\bibitem[Mulhern(2021)]{ai-environment}
Owen Mulhern.
\newblock Artificial intelligence: Can it help achieve environmental
  sustainability?
\newblock
  \url{https://earth.org/data_visualization/ai-can-it-help-achieve-environmental-sustainable/},
  2021.

\bibitem[{United Nation}(2021)]{block-chain-footprint}
{United Nation}.
\newblock {The Good, The Bad And The Blockchain}.
\newblock \url{https://unfccc.int/blog/the-good-the-bad-and-the-blockchain},
  2021.

\bibitem[Vollset et~al.(2020)Vollset, Goren, Yuan, Cao, Smith, Hsiao,
  Bisignano, Azhar, Castro, Chalek, Dolgert, Frank, Fukutaki, Hay, Lozano,
  Mokdad, Nandakumar, Pierce, Pletcher, Robalik, Steuben, Wunrow, Zlavog, and
  Murray]{Vollset}
SE~Vollset, E.~Goren, CW~Yuan, J~Cao, AE~Smith, T~Hsiao, C~Bisignano, GS~Azhar,
  E~Castro, J~Chalek, AJ~Dolgert, T~Frank, K~Fukutaki, SI~Hay, R~Lozano,
  AH~Mokdad, V~Nandakumar, M~Pierce, M~Pletcher, T~Robalik, KM~Steuben,
  HY~Wunrow, BS~Zlavog, and CJL Murray.
\newblock {Fertility, mortality, migration, and population scenarios for 195
  countries and territories from 2017 to 2100: a forecasting analysis for the
  Global Burden of Disease Study}.
\newblock \emph{Lancet}, 396, October 2020.

\bibitem[Roser(2019)]{UNDemographics}
Max Roser.
\newblock {Future Population Growth}.
\newblock \url{https://ourworldindata.org/future-population-growth}, November
  2019.

\bibitem[Jacobs(2015)]{Jacobs2015TheEO}
M.~Jacobs.
\newblock The effectiveness of word prediction software wordq: "…predict it,
  hear it, choose it, review it, correct it, write it now…".
\newblock 2015.

\bibitem[Haselton(2019)]{airpods}
Todd Haselton.
\newblock {This hidden AirPods feature helps you hear better}.
\newblock
  \url{https://www.cnbc.com/2019/01/20/how-to-turn-airpods-into-hearing-aids.html},
  2019.

\bibitem[Nachmani et~al.(2020)Nachmani, Adi, and Wolf]{nachmani2020voice}
Eliya Nachmani, Yossi Adi, and Lior Wolf.
\newblock Voice separation with an unknown number of multiple speakers, 2020.

\bibitem[Fearn(2020)]{vr}
Nicholas Fearn.
\newblock {How VR Is Helping Visually Impaired Patients Regain Close To Normal
  Levels Of Sight}.
\newblock
  \url{https://www.forbes.com/sites/nicholasfearn/2020/01/08/how-vr-is-helping-visually-impaired-patients-regain-close-to-normal-levels-of-sight/?sh=15e1a9c60277},
  2020.

\bibitem[Ghosh et~al.(2016)Ghosh, Donoghue, Khayrullin, Ali, Wacyk, Tice,
  Vazan, Sziklas, Fellowes, and Draper]{oled}
Amalkumar Ghosh, Evan~P. Donoghue, Ilyas Khayrullin, Tariq Ali, Ihor Wacyk,
  Kerry Tice, Fridrich Vazan, Laurie Sziklas, David Fellowes, and Russell
  Draper.
\newblock Invited paper: Directly patterened 2645 ppi full color oled
  microdisplay for head mounted wearables.
\newblock \emph{Society for Information Display Symposium Digest of Technical
  Papers}, 47\penalty0 (1), 2016.

\bibitem[Kubota(2020)]{oled-1}
Taylor Kubota.
\newblock {Stanford materials scientists borrow solar panel tech to create new
  ultrahigh-res OLED display}.
\newblock
  \url{https://news.stanford.edu/2020/10/22/future-vr-employ-new-ultrahigh-res-display/},
  2020.

\bibitem[Kodukula et~al.(2021)Kodukula, Katrawala, Jones, Wu, and
  LiKamWa]{Stagioni}
Venkatesh Kodukula, Saad Katrawala, Britton Jones, Carole-Jean Wu, and Robert
  LiKamWa.
\newblock Dynamic temperature management of near-sensor processing for
  energy-efficient high-fidelity imaging.
\newblock \emph{Sensors}, 21\penalty0 (3), 2021.

\bibitem[MacKenzie and Ho(2015)]{flexible-e}
J.~Devin MacKenzie and Christine Ho.
\newblock Perspectives on energy storage for flexible electronic systems.
\newblock \emph{Proceedings of the IEEE}, 103\penalty0 (4):\penalty0 535--553,
  2015.

\bibitem[Brunner et~al.(2011)Brunner, Bianchi, Guger, Cincotti, and
  Schalk]{BCI-2}
P~Brunner, L~Bianchi, C~Guger, F~Cincotti, and G.~Schalk.
\newblock Current trends in hardware and software for brain-computer interfaces
  (bcis).
\newblock In \emph{Journal of Neural Engineering}, 2011.

\bibitem[Karageorgos et~al.(2020)Karageorgos, Sriram, Veselý, Wu, Powell,
  Borton, Manohar, and Bhattacharjee]{BCI}
Ioannis Karageorgos, Karthik Sriram, Ján Veselý, Michael Wu, Marc Powell,
  David Borton, Rajit Manohar, and Abhishek Bhattacharjee.
\newblock Hardware-software co-design for brain-computer interfaces.
\newblock In \emph{Proceedings of the ACM/IEEE Annual International Symposium
  on Computer Architecture}, 2020.

\bibitem[{International Energy Agency}()]{world-electricity}
{International Energy Agency}.
\newblock {Access to electricity}.
\newblock
  \url{https://www.iea.org/reports/sdg7-data-and-projections/access-to-electricity}.

\bibitem[Isaacman and Martonosi(2011)]{CLink}
Sibren Isaacman and Margaret Martonosi.
\newblock Low-infrastructure methods to improve internet access for mobile
  users in emerging regions.
\newblock In \emph{Proceedings of the 20th International Conference Companion
  on World Wide Web}, page 473–482, 2011.

\bibitem[Hester et~al.(2013)Hester, King, Propst, Piratla, and Sorber]{poster}
Josiah Hester, Trae King, Alex Propst, Kalyan Piratla, and Jacob Sorber.
\newblock Enabling sustainable sensing in adverse environments.
\newblock In \emph{Proceedings of the IEEE International Conference on Sensing,
  Communications and Networking}, pages 249--251, 2013.

\bibitem[Lucia et~al.(2017)Lucia, Balaji, Colin, Maeng, and
  Ruppel]{intermittent-computing}
Brandon Lucia, Vignesh Balaji, A.~Colin, Kiwan Maeng, and E.~Ruppel.
\newblock Intermittent computing: Challenges and opportunities.
\newblock In \emph{SNAPL}, 2017.

\bibitem[Saleem et~al.(2020)Saleem, Schmitt, Chen, and
  Raghavan]{demographic-inclusion-2}
Bilal Saleem, Paul Schmitt, Jay Chen, and Barath Raghavan.
\newblock Beyond the trees: Resilient multipath for last-mile {WISP} networks.
\newblock \emph{CoRR}, abs/2002.12473, 2020.

\bibitem[Yetim and Martonosi(2015)]{delay-tolerance-prediction}
Ozlem~Bilgir Yetim and Margaret Martonosi.
\newblock Dynamic adaptive techniques for learning application delay tolerance
  for mobile data offloading.
\newblock In \emph{2015 IEEE Conference on Computer Communications (INFOCOM)},
  pages 1885--1893, 2015.

\bibitem[Plumer(2021)]{energy-grid}
Brad Plumer.
\newblock {Energy Department Targets Vastly Cheaper Batteries to Clean Up the
  Grid}.
\newblock
  \url{https://www.nytimes.com/2021/07/14/climate/renewable-energy-batteries.html},
  2021.

\bibitem[Stray(2021)]{recsys}
Jonathan Stray.
\newblock {Beyond Engagement: Aligning Algorithmic Recommendations With
  Prosocial Goals}.
\newblock
  \url{https://www.partnershiponai.org/beyond-engagement-aligning-algorithmic-recommendations-with-prosocial-goals/},
  2021.

\bibitem[Askell et~al.(2019)Askell, Brundage, and Hadfield]{openai-rai}
Amanda Askell, Miles Brundage, and Gillian Hadfield.
\newblock The role of cooperation in responsible ai development, 2019.

\bibitem[{NIST}(2021)]{nist-trustworthy}
{NIST}.
\newblock Nist proposes approach for reducing risk of bias in artificial
  intelligence.
\newblock
  \url{https://www.nist.gov/news-events/news/2021/06/nist-proposes-approach-reducing-risk-bias-artificial-intelligence},
  2021.

\bibitem[IBM()]{ibm-rai}
IBM.
\newblock Ai ethics.
\newblock \url{https://www.ibm.com/artificial-intelligence/ethics}.

\bibitem[Microsoft({\natexlab{b}})]{microsoft-rai}
Microsoft.
\newblock {Trustworthy AI}.
\newblock
  \url{https://www.microsoft.com/en-us/research/project/trustworthy-ai/},
  {\natexlab{b}}.

\bibitem[Google({\natexlab{c}})]{google-rai}
Google.
\newblock Building responsible ai for everyone.
\newblock \url{https://ai.google/responsibilities/}, {\natexlab{c}}.

\bibitem[Facebook()]{facebook-rai}
Facebook.
\newblock {Facebook’s five pillars of Responsible AI}.
\newblock
  \url{https://ai.facebook.com/blog/facebooks-five-pillars-of-responsible-ai/}.

\bibitem[DOD()]{dod-rai}
DOD.
\newblock {AI Principles: Recommendations on the Ethical Use of Artificial
  Intelligence by the Department of Defense}.
\newblock
  \url{https://media.defense.gov/2019/Oct/31/2002204458/-1/-1/0/DIB_AI_PRINCIPLES_PRIMARY_DOCUMENT.PDF}.

\bibitem[Roser(2020)]{a4ai}
Max Roser.
\newblock {Mobile devices are too expensive for billions of people -- and it's
  keeping them offline}.
\newblock
  \url{https://a4ai.org/mobile-devices-are-too-expensive-for-billions-of-people-and-its-keeping-them-offline/},
  August 2020.

\bibitem[Naseem et~al.(2020)Naseem, Saleem, St-Onge~Ahmad, Chen, and
  Raza]{demographic-inclusion-1}
Mustafa Naseem, Bilal Saleem, Sacha St-Onge~Ahmad, Jay Chen, and Agha~Ali Raza.
\newblock \emph{An Empirical Comparison of Technologically Mediated Advertising
  in Under-Connected Populations}, page 1–13.
\newblock 2020.

\bibitem[O’Neill et~al.(2018)O’Neill, Fanning, Lamb, and
  Steinberger]{ONeill2018AGL}
Daniel~W. O’Neill, Andrew~L. Fanning, William~F. Lamb, and J.~Steinberger.
\newblock A good life for all within planetary boundaries.
\newblock \emph{Nature Sustainability}, 1:\penalty0 88--95, 2018.

\end{thebibliography}
\end{flushleft}






\end{document}